\documentclass[%
 reprint,
 superscriptaddress,
 amsmath,amssymb,
 aps,
 pre,
]{revtex4-2}

\usepackage{comment}
\usepackage{graphicx}% Include figure files
\usepackage{dcolumn}% Align table columns on decimal point
\usepackage{bm}% bold math
\usepackage{xcolor}
\usepackage{pifont}
\usepackage{hyperref}% add hypertext capabilities
\usepackage{xr}
\def\kt{k_\text{B}T}
\newcommand*{\Kd}{K_\text{c}}
\newcommand*{\fc}{f_\text{c}}
\newcommand*{\rhoT}{\rho_{\text{T}}}
\newcommand*{\rhoB}{\rho_1^\text{bg}}
\newcommand*{\rhoC}{\rho_1^\text{c}}

\newcommand*{\Vr}{V_\text{r}}

\newcommand*{\VT}{V_\text{tot}}

\newcommand{\Econd}{\epsilon_{\text{c}}}
\newcommand{\ebind}{\epsilon_{\text{ss}}}
\newcommand{\Rc}{R_{\text{c}}}
\newcommand{\Ucond}{u_{\text{c}}}
\newcommand{\alphac}{\alpha_{\text{C}}}

\newcommand{\Kc}{K_{\text{c}}}

\newcommand{\Vc}{V_{\text{c}}}
\newcommand{\Vbg}{V_{\text{bg}}}

\newcommand{\rhoonebg}{\rho_1^{\text{bg}}}
\newcommand{\rhoonec}{\rho_1^{\text{c}}}

\newcommand{\fcp}{f_{\text{CP}}}
\newcommand{\rhocp}{\rho_{\text{CP}}}

\newcommand{\sigcap}{\sigma_{\text{cap}}}
\newcommand{\tHalf}{\tau_{1/2}}
\newcommand{\rMax}{r_\text{max}}

\newcommand{\KcIS}{\Kc^\text{IS}}

\newcommand{\ncap}{N_{\text{cap}}}
\newcommand{\rhonc}{\rho_{\text{cap}}^{\text{c}}}

\newcommand{\rhoStar}{\rho_\text{CSC}}

\newcommand{\rhon}{\rho_{\text{cap}}}

\newcommand*{\sNuc}{s_\text{nuc}}

\newcommand*{\nNuc}{{n_\text{nuc}}}

\newcommand*{\tF}{t_\text{F}}
\newcommand{\KcActual}{\Kc^\text{meas}}
\newcommand{\tDiff}{\tau_\text{D}}
\newcommand{\tHalfMin}{\tau_{1/2}^\text{min}}

\begin{document}

\title{Computer simulations show that liquid-liquid phase separation enhances self-assembly}

\author{Layne B. Frechette}
\thanks{These two authors contributed equally.}
\affiliation{
 Martin Fisher School of Physics, Brandeis University, Waltham, Massachusetts 02453, USA
}

\author{Naren Sundararajan}
\thanks{These two authors contributed equally.}
\affiliation{
 Martin Fisher School of Physics, Brandeis University, Waltham, Massachusetts 02453, USA
}

\author{Fernando Caballero}
\affiliation{
 Martin Fisher School of Physics, Brandeis University, Waltham, Massachusetts 02453, USA
}
\author{Anthony Trubiano}
\affiliation{
 Martin Fisher School of Physics, Brandeis University, Waltham, Massachusetts 02453, USA
}
\author{Michael F. Hagan}
\email{hagan@brandeis.edu}
\affiliation{
 Martin Fisher School of Physics, Brandeis University, Waltham, Massachusetts 02453, USA
}

\date{\today}
\begin{abstract}
Biomolecular condensates are liquid- or gel-like droplets of proteins and nucleic acids formed at least in part through liquid-liquid phase separation. Condensates enable diverse functions of cells and the pathogens that infect them, including self-assembly reactions. For example, it has been shown that many viruses form condensates within their host cells to compartmentalize capsid assembly and packaging of the viral genome.  Yet, the physical principles controlling condensate-mediated self-assembly remain incompletely understood. In this article we use coarse-grained molecular dynamics simulations to study the effect of a condensate on the assembly of icosahedral capsids. The capsid subunits are represented by simple shape-based models to enable simulating a wide range of length and time scales, while the condensate is modeled implicitly to study the effects of phase separation independent of the molecular details of biomolecular condensates. Our results show that condensates can significantly enhance assembly rates, yields, and robustness to parameter variations, consistent with previous theoretical predictions. However, extending beyond those predictions, the computational models also show that excluded volume enables control over the number of capsids that assemble within condensates. Moreover, long-lived aberrant off-pathway assembly intermediates can suppress yields within condensates. In addition to elucidating condensate-mediated assembly of viruses and other biological structures, these results may guide the use of condensates as a generic route to enhance and control self-assembly in human-engineered systems. 
\end{abstract}
\maketitle

\section{Introduction}\label{sec:intro}

Many critical biological functions rely on high-fidelity assembly of structures, such as viral capsids \cite{Caspar1962, Zlotnick2011,Mateu2013,Bruinsma2015,Perlmutter2015,Hagan2016,Twarock2018,Zandi2020}, 
photonic nanostructures \cite{R.Dufresne2009}, bacterial microcompartments \cite{Abeysinghe2024, Chowdhury2014,Kerfeld2010, Kerfeld2015, Kerfeld2016, Erbilgin2014, Price1991, Shively1973}, and other proteinaceous organelles \cite{Sutter2008,Nichols2020,Pfeifer2012,Kickhoefer1998,Zaslavsky2018}. These structures form with astonishing robustness, despite the complex and highly adaptive nature of the cellular cytoplasm. 
Self-assembly also has promising technological applications, such as enabling highly scalable bottom-up manufacturing of nanostructured materials~\cite{Li2022,Fendler1996,Vanmaekelbergh2011}. Recent developments in DNA origami, protein design, supramolecular assembly, and patchy-colloidal particles have enabled the design of human-engineered subunits that assemble into complex architectures including icosahedral shells and cylindrical tubules with programmable sizes (e.g. ~\cite{Rothemund2006, Benson2015, Sigl2021, Videbaek2022, Hayakawa2022, Wei2024, Videbaek2024, Wagenbauer2017, Divine2021, Butterfield2017, Bale2016, Hsia2016, King2012, Dowling2023, Malay2019, Ren2019, Laniado2021, McConnell2020, McMullen2022, Garg2015, Beija2012, Ebbens2016, Mallory2018, Fan2011, Huh2020, Ke2012, J.Kraft2009, Sacanna2010, Sacanna2013, Yi2013, Wang2014, He2020, He2021, Chen2011a, Chen2011b, Zerrouki2008, Yan2013, R.Wolters2015, Tikhomirov2018, Oh2019, Ben-Ari2021, Kahn2022}). However, in contrast to biology, the ability to achieve high-fidelity assembly in these synthetic systems is far from robust, requiring precise tuning of subunit concentrations and interactions to avoid kinetic traps (long-lived off-pathway intermediates). \cite{Wei2024, Jacobs2025, Zandi2020, Hagan2006, Wilber2007, Whitelam2015, Panahandeh2020, Nguyen2007, Zlotnick1999, Endres2002, Zlotnick2003, Ceres2002, Jack2007, Rapaport2008, Whitelam2009, Wilber2009, Hagan2011, Cheng2012, Hagan2014, Perlmutter2016, Asor2020, Qian2023}. Biological systems employ multiple mechanisms to avoid such kinetic traps. In this article we use computer simulations to investigate one of these mechanisms --- self-assembly coupled to biomolecular condensates. Our results show that condensates can significantly increase assembly rates and robustness, and enable controlling the final yield. 

It has become clear that biomolecular condensates formed through liquid-liquid phase separation (LLPS) play a key role in spatially organizing cellular environments \cite{Julicher2024, Hyman2012, Berry2018, Choi2020, Banani2017}. These `membraneless organelles' are involved in essential processes, such as transcriptional regulation \cite{Cho2018, Henninger2021, Hnisz2017, Sabari2018, Chong2018}, cell division \cite{Parker2007, Brangwynne2009}, and neuronal synapse formation \cite{Wu2020, Stanishneva-Konovalova2016, DelSignore2021}. More recent evidence suggests that condensates can also control self-assembly within cells. Examples include the assembly of clathrin cages during endocytosis \cite{Day2021}, the formation of post-synaptic densities \cite{Zeng2018a} and pre-synaptic vesicle release sites \cite{McDonald2020, Emperador-Melero2021} at neuronal synapses, aggregation of $\alpha$-synuclein \cite{Lipinski2022}, and actin assembly in  polypeptide coacervates \cite{McCall2018}. In addition to normal biological functions, condensates are also critical for pathogenic infections. For example, in many viral infections, the outer protein shells (capsids) assemble within condensates known as virus factories, replication sites, Negri bodies, inclusion bodies, or viroplasms \cite{Borodavka2018,Borodavka2017,Brocca2020,Carlson2020,Etibor2021,FernandezdeCastro2020,Gaete-Argel2019,Geiger2021,Guseva2020,Kieser2020,Lopez2021,Luque2020,Nikolic2017,Pan2021,Papa2021,Risso-Ballester2021,Savastano2020,Schoelz2017,Trask2012,Wang2021}. However, the effects of LLPS on assembly and the benefits to the viral lifecycle remain incompletely understood.

\begin{figure}[ht]
    \centering
    \includegraphics[width=\linewidth]{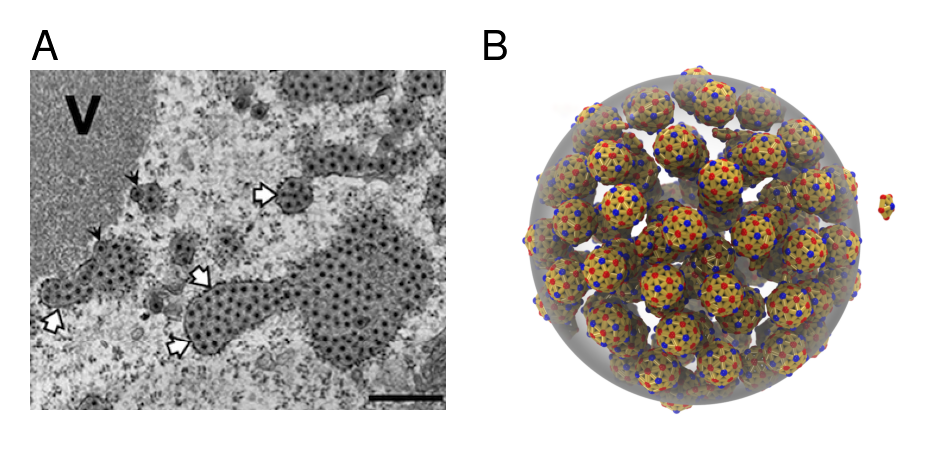}
    \caption{Densely-packed viral factories in experiments and computer simulations. (A) Electron microscopy image from a rotavirus-infected cell. Gray regions are viroplasms, and black dots are assembled capsids. V, viroplasm; black arrowheads, enveloped viral particles; white arrows, endoplasmic reticulum surrounding viroplasms. Adapted from Papa et al. ~\cite{Papa2019}, which is licensed under \href{https://creativecommons.org/licenses/by/4.0/}{CC BY 4.0}. (B) Snapshot from a computer simulation of LLPS-coupled capsid assembly, showing capsids (small spherical particles) densely packed within a condensate (large gray sphere). 
    }
    \label{fig:sim_vs_expt}
\end{figure}

Extensive theoretical and computational works have investigated how condensates form and how they are controlled by chemical reactions and other nonequilibrium processes (e.g. \cite{Zwicker2014a, Zwicker2015, Zwicker2017, Zwicker2025, Weber2019c, Ziethen2023, Ziethen2024, Soding2020, Kirschbaum2021, Luo2025, Rossetto2024, Choi2019, Brangwynne2015, Ruff2019, Holla2024, Dignon2018c, Dignon2019a, Dignon2019b, Shea2021, Rekhi2024, Kapoor2024, Hnisz2017, Sabari2018, Schede2023, Banani2024,Pyo2023,GrandPre2023,Zhang2024,Grigorev2025,Jacobs2021,Chen2024,Li2024,Lin2018,Cinar2019,Pal2021,Shen2023,Das2018,Nguemaha2018,Kota2022,Zhou2024}). However, there have been comparatively few studies of condensate-coupled assembly. Recent investigations using chemical kinetics-based rate equations suggest that preferential partitioning of subunits into condensates can significantly enhance assembly rates by locally concentrating the subunits \cite{Bartolucci2024, Hagan2023}. Ref. \cite{Hagan2023} also showed that LLPS can make assembly more robust to parameter variations, significantly broadening the range of subunit concentrations or binding affinities that lead to productive assembly. Although Bartolucci et al. \cite{Bartolucci2024} made a comprehensive study that included effects of subunits on condensate phase coexistence, in many systems the subunits remain sufficiently dilute within condensates so that their effects on condensate stability can be neglected 
(i.e., subunits are ``clients'' rather than ``scaffolds''~\cite{Ditlev2018}). Despite the important insights from these works, they were limited to rate equation models that assume spatially uniform concentrations within condensates. Further, they must pre-assume the set of allowed assembly intermediates and thus have not considered the excluded volume geometries of assembly structures or the possibility of malformed structures and other off-pathway intermediates that can limit assembly of target structures (e.g. \cite{Perlmutter2015, Mateu2013, Zandi2020, Whitelam2015, Hagan2014, Mohajerani2022, Sigl2021, Hagan2021, Li2018, Perlmutter2015, Mateu2013, Zandi2020, Schwartz2000,Rapaport2004,Nguyen2009,Elrad2008,Johnston2010,Rapaport2010,Nguyen2008, Schwartz1998,Sweeney2008, Panahandeh2020, Hagan2006,Jack2007,Rapaport2008,Whitelam2009,Nguyen2007,Wilber2007,Wilber2009,Hagan2011,Cheng2012,Tyukodi2022,Fang2022, Rotskoff2018}). 

In this article, we use molecular dynamics simulations to avoid these approximations. To enable conclusions that are independent of the molecular details of condensates and to simulate a wide range of lengths and timescales, we implicitly model the condensate. That is, we represent the condensate as a spherical region with an attractive field that drives partitioning of subunits into the condensate. The simulation results are consistent with many of the predictions of the rate equation models, such as enhanced yields, rates and robustness to parameter variations, but also identify behaviors that are not captured in rate equation models. For example, capsid yields are suppressed by kinetic traps due to aberrant off-pathway structures for strong subunit binding affinities. Capsid yields also drop precipitously when the concentration of subunits approaches the close-packing limit. We develop an equilibrium theory that recapitulates this behavior, indicating that this reduction in yield is at least partially thermodynamic in origin. In this limit, we observe para-crystalline arrays of nearly closely packed capsids, which are highly reminiscent of recent electron microscopy images of virus factories (Fig.~\ref{fig:sim_vs_expt}). We show that capsid yields can be precisely controlled by the condensate volume. In addition to shedding light on viral lifecycles, this suggests a strategy to achieve size-controlled crystalline arrays in synthetic assembly systems.

\begin{figure*}[ht]
    \centering
    \includegraphics[width=\linewidth]{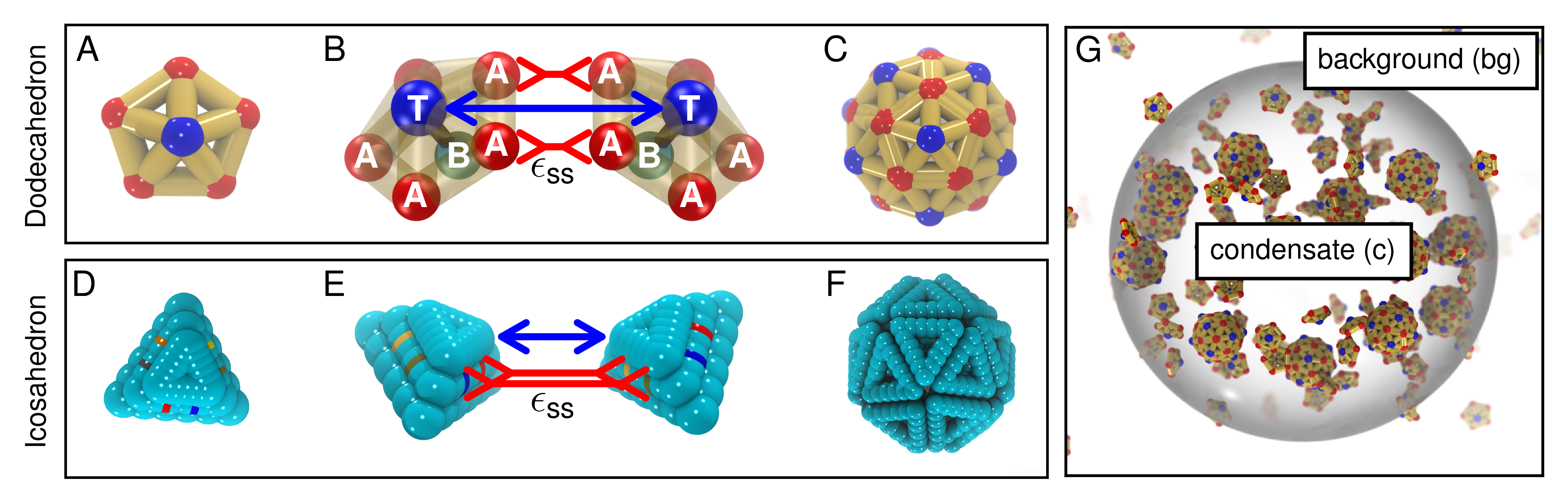}
    \caption{Illustration of the capsid models. (A) Dodecahedron model subunit. (B) Interactions between dodecahedron subunits. Top (`T') pseudoatoms repel each other, while attractor (`A') pseudoatoms attract each other with interaction strength $\ebind$. Bottom (`B') pseudoatoms repel `T' pseudoatoms, helping to prevent misbinding. (C) Dodecahedral capsid.  (D) Icosahedron model subunit. (E) Interactions between icosahedron subunits. Excluder pseudoatoms (cyan) repel each other, while attractor pseudoatoms (all other colors) attract each other with interaction strength $\ebind$. (F) Icosahedral capsid. (G) Snapshot of a condensate containing  monomeric subunits and capsids, taken from an assembly simulation of the dodecahedron model. Subunits can exchange between the condensate (gray sphere) with volume $\Vc$ and the background with volume $\Vbg$. At equilibrium, the concentrations of monomers in the condensate and background are related by the partition coefficient, $\Kc=\rhoonec/\rhoonebg$.
    }
    \label{fig:model}
\end{figure*}

\section{Model and Methods}\label{sec:methods}
We use molecular dynamics to simulate  $N$ subunits in a cubic box with edge-length $L$ and volume $V = L^3$ (subject to periodic boundary conditions) at a temperature $T$. We study assembly across a range of total subunit concentrations $\rhoT=N/V$. We perform our simulations using the GPU-accelerated HOOMD-blue package \cite{Anderson2020} (v2.9.0 for the icosahedron model, v4.6.0 and v4.8.0 for the dodecahedron model).

To assess the generality of our results, we study two capsid models, which differ in complexity and in the size of the assembled structures. To reduce the computational cost of simulating high-frequency vibrational motions, both models represent capsid subunits as rigid bodies. In the first model (Fig.~\ref{fig:model}A-C), inspired by SV40 virus capsids~\cite{Perlmutter2013,Perlmutter2014} pentagonal subunits (Fig.~\ref{fig:model}A) with circumradius $l_0$ assemble into dodecahedral capsids (Fig.~\ref{fig:model}C). The subunits consist of a top (`T') pseudoatom that has repulsive interactions with other `T' pseudoatoms, a bottom (`B') pseudoatom that has repulsive interactions with `T' pseudoatoms to help prevent subunits from binding with upside-down configurations, and five attractor (`A') pseudoatoms located at the vertices of the pentagon that account for subunit-subunit attractions ((Fig.~\ref{fig:model}B and SI Section S1). The `T' and `B' interactions are represented by a WCA-like potential \cite{Weeks1971}. The `A' pseudoatoms interact via a Morse potential, for which the well-depth parameter $\ebind$ sets the subunit-subunit binding affinity, which we estimate in SI Section S4.

The second, more complex, model (Fig.~\ref{fig:model}D-F)  is inspired by DNA origami assembly, and consists of triangular subunits (Fig.~\ref{fig:model}D) of diameter $\approx 3\sigma$ (where $\sigma$ is the diameter of the pseudoatoms comprising each subunit) that assemble into $T=1$ icosahedral capsids (Fig.~\ref{fig:model}F)~\cite{Wei2024}. These subunits consist of 45 excluder pseudoatoms that account for excluded volume via a WCA potential, and six attractor pseudoatoms which bind to complementary attractor atoms via a Lennard-Jones potential with well depth $\ebind$ (see Fig.~\ref{fig:model}B and SI Section S1). We describe the models in detail in the Supporting Information (SI) Section S1.

\subsection{Condensate model}\label{sec:condensate_model}

We model the condensate as a spherical region of radius $\Rc$ centered at the origin. The condensate imposes a spherically symmetric potential given by:
\begin{equation}
    \Ucond(r) = \begin{cases} -\Econd, & r<\Rc \\
        -\Econd\left(2e^{-2\alphac(r-\Rc)}-e^{-2\alphac(r-\Rc)}\right), & r\geq \Rc
    \end{cases} \label{eq:ucond}
\end{equation}
where $r$ is the distance from the origin, $\Econd$ sets the depth of the potential well, and $\alphac$ is a parameter that controls how rapidly the potential decays to zero as $r$ increases beyond $\Rc$. We set $\alphac=$ 10 $l_0^{-1}$ ($10\sigma^{-1}$ for the icosahedron model), which ensures that the potential goes to zero over a length scale $\alphac^{-1}\ll \Rc$ while suppressing unphysically large forces at the condensate boundary. The condensate occupies a volume $\Vc$, and the background has volume $\Vbg=V-\Vc$. We define the condensate volume ratio as $\Vr=\Vc/\Vbg$ to characterize the relative sizes of the condensate and background independently of the total volume. Except where otherwise noted, we use $\Vr=5.03\times 10^{-3}$ for the dodecahedron model and $\Vr=4.20\times 10^{-3}$ for the icosahedron model. These values are in the typical range for condensates within eukaryotic cells \cite{Banani2017}. 

The well-depth $\Econd$ controls the strength of subunit-condensate interactions. For $\Econd>0$, subunits preferentially partition into the condensate, as characterized by the partition coefficient:
\begin{equation}
    \Kc = \frac{\rho_1^{\text{c}}}{\rho_1^{\text{bg}}},
\end{equation}
where $\rho_1^{\text{c}}, \rho_1^{\text{bg}}$ are the equilibrium concentrations of unassembled subunits in the condensate and background (volume outside the condensate). In the ideal solution (IS) limit (for packing fractions $\eta = \frac{\pi}{6}\rho l_0^3 \lesssim 0.1$) the partition coefficient is related to the well depth by $\KcIS = e^{\beta \Econd}$, and assembly will significantly increase partitioning of larger structures, since the driving force for a cluster with $n$ subunits to partition into the condensate is $\propto n \Econd$. However, for higher concentrations excluded volume effects will reduce the partition coefficient, particularly once capsids assemble due to their large excluded volume. We account for excluded  volume effects of monomers alone (i.e. before assembly occurs) as well as in the presence of assembly by modeling subunits and capsids as effective hard spheres using the
 the Carnahan-Starling equation of state \cite{Carnahan1969,Carnahan1970} and its extension to binary mixtures \cite{Mansoori1971} (see Figs.~\ref{fig:Kc_vs_eps_c}, S6, and SI Section S2). See the Appendix for additional discussion of how $\Kc$ depends on $\Econd$. A snapshot of a condensate containing both monomeric subunits and capsids is shown in Fig.~\ref{fig:model}G.

\subsection{Units}
\label{sec:units}
In both the dodecahedron and icosahedron models, the thermal energy $\kt$ (where $k_{\text{B}}$ is Boltzmann's constant) serves as our unit of energy and the subunit mass $m$ is the unit of mass. In the dodecahedron model, $l_0$ is the unit length, and $t_0\equiv l_0 \sqrt{m/\kt}$ is the unit time. For the icosahedron model, $\sigma$ is the unit length and $t_0\equiv\sigma \sqrt{m/\kt}$ is the unit time. We report all quantities in terms of these units (energies in $\kt$, lengths in $l_0$ for the dodecahedron model and $\sigma$ for the icosahedron model, and times in $t_0$).

\subsection{Simulation details}

For the dodecahedron model, we use $N=1200$ subunits, and vary the box size from $L=288.5$ to $106.3$ to control the concentration over the range $\rhoT \in [5\times 10^{-5},10^{-3}]$. For the icosahedron model, we use  $L=200$ and $N=149$ to $2092$ to achieve $\rhoT \in [1.8\times 10^{-5},2.6\times10^{-4}]$.

Trajectories are initialized with subunits placed in the box at random, non-overlapping positions. We then run Langevin dynamics for $\tF=10^6$ ($1.2\times 10^8$ time steps  with time step $\Delta t=5\times 10^{-3}$ for the dodecahedron model and  $2\times 10^8$ time steps  with $\Delta t=5\times 10^{-3}$ for the icosahedron model) unless noted otherwise. All reported results were obtained by averaging over 10 (dodecahedron) or 5 (icosahedron) independent trajectories for each parameter set ($\ebind$, $\Econd$, and $\rhoT$). Error bars represent twice the standard error of the mean (approximately a 95\% confidence interval).

We used our open-source Python library SAASH \cite{Trubiano2024} to analyze trajectories and calculate the yield of complete capsids. The library provides functions to identify fully assembled capsids as well as arbitrarily-sized clusters. For cluster identification, we consider two dodecahedron subunits to be bonded (and hence part of the same cluster) if the centers of the edges of two subunits are within $0.3$ of each other; we consider two icosahedron subunits to be bonded if both pairs of attractor pseudoatoms on the two subunits are within $1.3$ of each other. Additional Python scripts used in this work have been deposited on Github at the following URL: https://github.com/Layne28/implicit-condensate.

\section{Results}

\subsection{LLPS promotes robust assembly}\label{sec:robust}

\begin{figure}
    \centering   
    \includegraphics[width=1.0\linewidth]{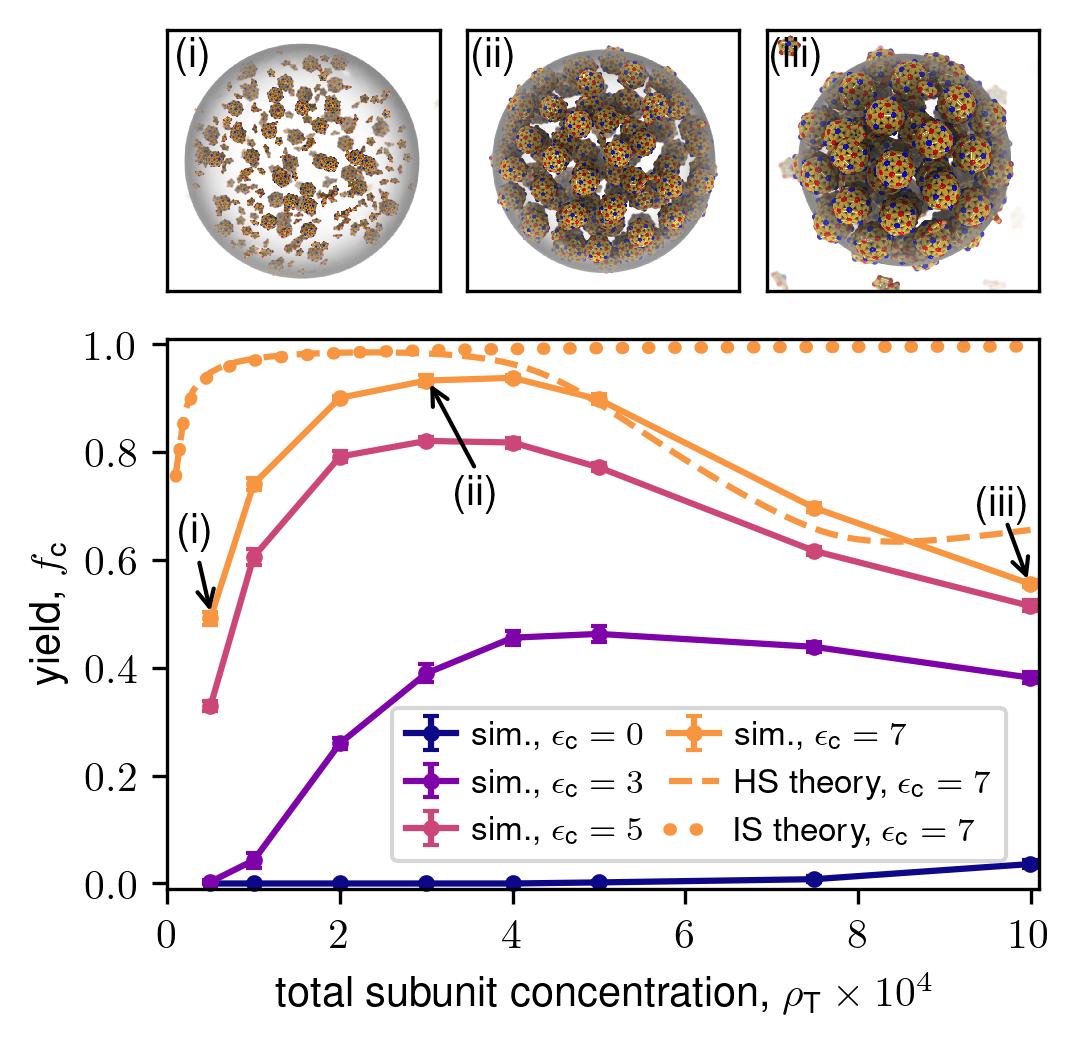}
    \caption{Final capsid yield versus total subunit concentration $\rhoT$ for the dodecahedron model for indicated values of $\Econd$, which controls the partition coefficient $\Kc$. Finite-time simulation yields are shown as symbols with lines. The  hard sphere (HS) theory (dashed line) and ideal solution (IS) theory (dotted line) are shown for $\Econd=7$. Markers labeled (i-iii) correspond to snapshots above the plot. Parameters are: the subunit-subunit binding energy $\ebind =6$, condensate volume fraction $\Vr=5.0\times 10^{-3}$, and final simulation time $\tF=6\times10^5$. Throughout this article, results are presented in dimensionless units defined in section~\ref{sec:units}. Except for Fig.~\ref{fig:yield_vs_concentration_icosahedron}, results are shown for the dodecahedron model.
    }
    \label{fig:yield_vs_concentration_dodecahedron}
\end{figure}

\begin{figure}
    \centering
    \includegraphics[width=1.0\linewidth]{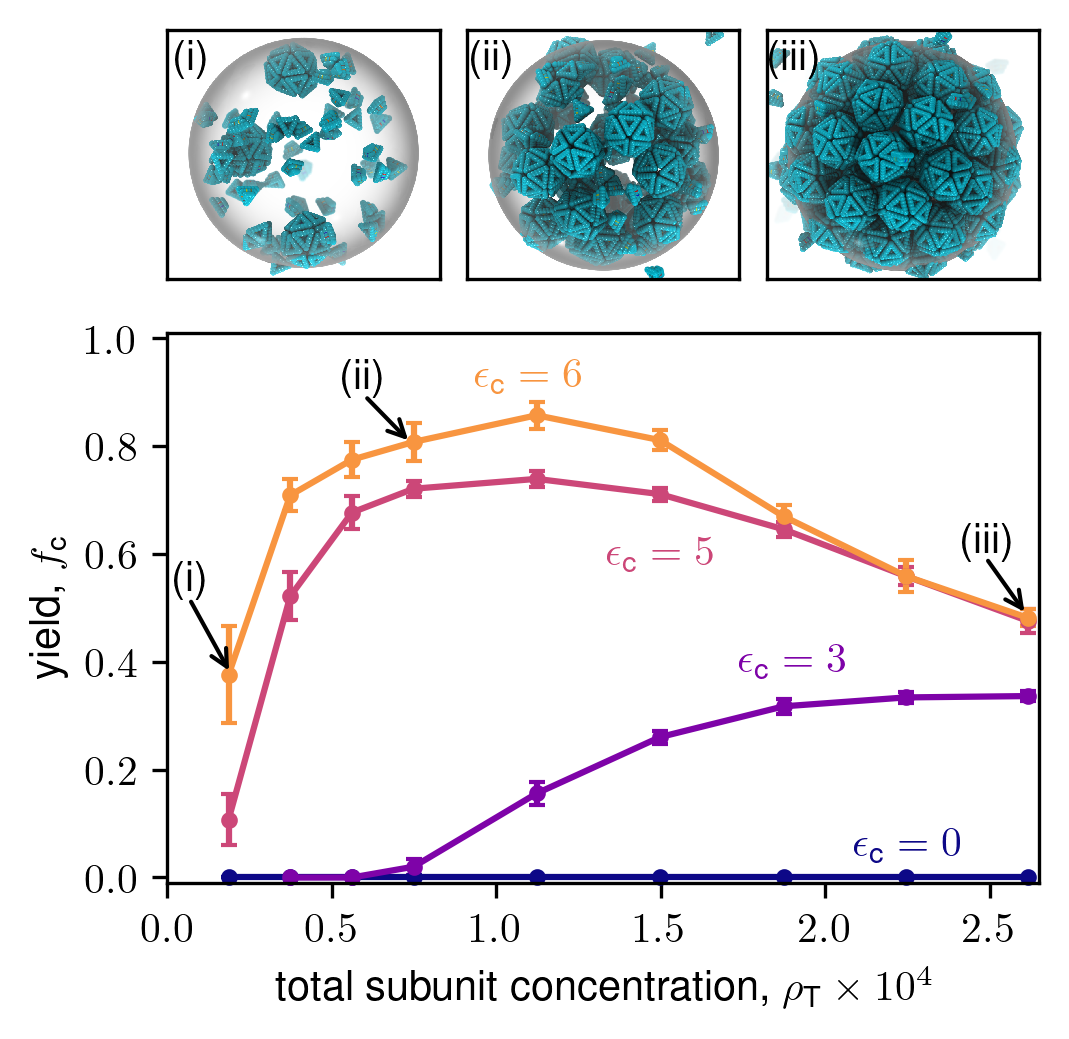}
    \caption{Computational results for final capsid yield versus total subunit concentration for the icosahedron model for indicated values of $\Econd$. Markers labeled (i-iii) correspond to snapshots above the plot. Parameters are: $\ebind=6$, $\Vr = 4.20 \times 10^{-3}$, and $\tF=10^6$. 
    }
    \label{fig:yield_vs_concentration_icosahedron}
\end{figure}

\begin{figure}
    \centering
    \includegraphics[width=1.0\linewidth]{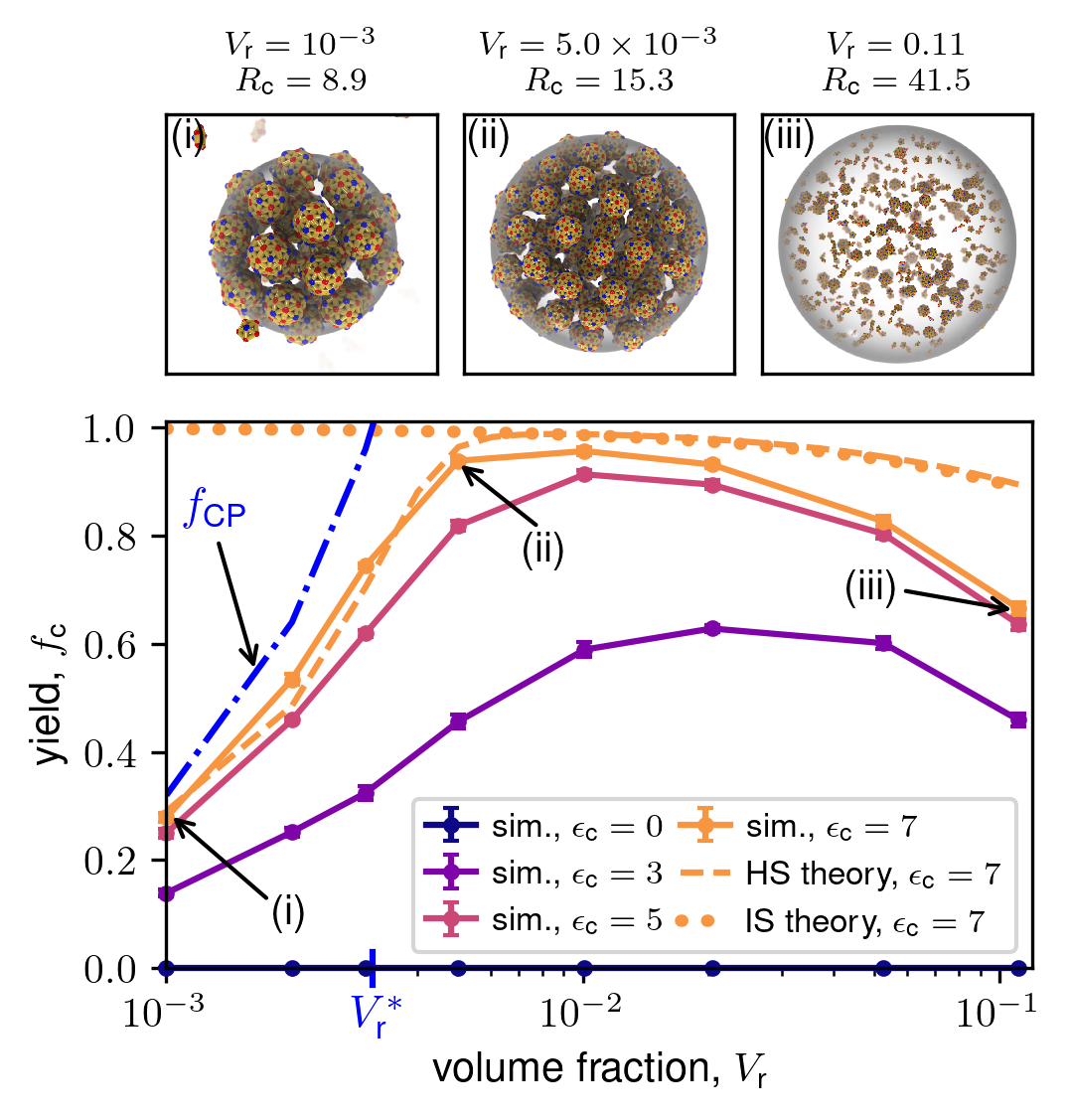}
    \caption{Capsid yield versus condensate volume fraction for indicated values of $\Econd$ for the dodecahedron model. Finite-time simulation yields are shown as symbols with lines; the hard sphere theory (HS, dashed line) and ideal solution theory (IS, dotted line) predictions are shown for $\Econd=7$. Points labeled with Roman numerals correspond to snapshots above. Supplementary Movie S5 shows the trajectory from which snapshot (iii) was taken. The value $\Vr^*$, marked in blue on the $x$ axis, is the threshold volume fraction below which the close-packed yield theory (see section~\ref{sec:excluded_volume}) predicts a decline in equilibrium yield. The estimated highest possible yield $\fcp$ (Eq.~\ref{eq:yield_fcp}, corresponding to spheres with approximately the size of capsids at close-packing density) is shown by the blue dash-dot line. Parameters are $\ebind =6$, $\rhoT=4.00\times 10^{-4}$, and $\tF=6\times10^5$.
}
    \label{fig:yield_vs_vr}
\end{figure}

\begin{figure*}[ht]
    \centering
    \includegraphics[width=1.0\linewidth]{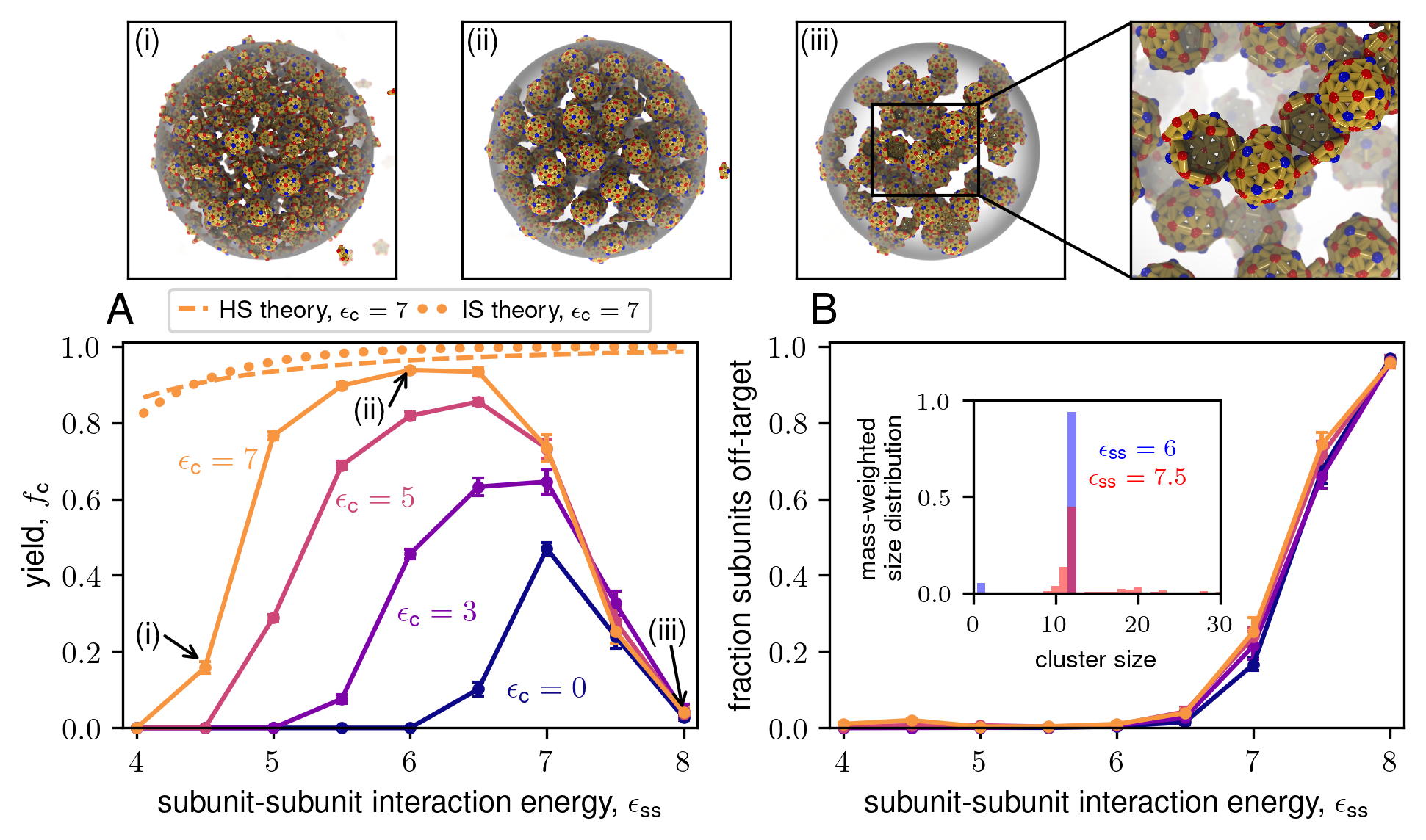}
    \caption{(A) Yield as a function of the subunit binding affinity parameter ($\ebind$) at indicated values of $\Econd$ for the dodecahedron model. Snapshots (i-iii) correspond to the labeled points at three different values of $\ebind$ with $\Econd=7$. 
The inset in (iii) shows an example of a large, malformed assembly consisting of several half-shells bound together. See Supplementary Movie S7 for the trajectory corresponding to snapshot (iii). (B) Fraction of subunits in off-target structures as a function of $\ebind$. The inset shows the size distributions at $\ebind=6$ and $\ebind=7.5$. Parameters for (A) and (B) are $\rhoT=4\times 10^{-4}$ and $\Vr=5.0\times 10^{-3}$. 
 } 
    \label{fig:yield_vs_ess}
\end{figure*}

In a uniform bulk solution, subunit concentration and binding affinity must be carefully tuned to achieve high yields (i.e. the fraction of subunits in complete capsids, $\fc$) \cite{Ceres2002,Katen2009,Hagan2014,Bruinsma2021}. Concentrations and binding affinities larger than their optimal values lead to kinetic traps, such as malformed assemblies or ``monomer starvation'' (in which many partial assemblies nucleate but do not form complete capsids due to rapid depletion of free monomers \cite{Ceres2002,Zlotnick2003,Hagan2006,Hagan2011}). On the other hand, concentrations and binding affinities that are too low result in nucleation times that exceed typical time scales of experiments or simulations. Our computational results illustrate these behaviors;  in the absence of phase separation ($\Econd=0$) we observe significant yields only in a narrow range of subunit-subunit interaction energies $\ebind$ (Figs.~\ref{fig:yield_vs_concentration_dodecahedron},~\ref{fig:yield_vs_concentration_icosahedron},~\ref{fig:yield_vs_vr},~\ref{fig:yield_vs_ess}, and Supplementary Movies S1, S2). We define yield as the fraction of subunits in complete capsids, where a complete capsid for the dodecahedron or icosahedron model has 12 or 20 subunits respectively, and each subunit forms the maximum number of bonds (as defined in Section~\ref{sec:methods}C) with its neighbors, 5 or 3 respectively. 

The computational results show that LLPS dramatically improves capsid assembly rates and yields in comparison to bulk assembly (Figs.~\ref{fig:yield_vs_concentration_dodecahedron},~\ref{fig:yield_vs_concentration_icosahedron},~\ref{fig:yield_vs_vr},~\ref{fig:yield_vs_ess} and Supplementary Movies S3, S4, S5). While these trends are qualitatively consistent with rate equation models for assembly coupled to LLPS developed in Ref.~\cite{Hagan2023}, the computational model also exhibits important differences resulting from capsid excluded volume and off-pathway intermediates, which are not accounted for in the previous theory. 

\subsubsection{There is an optimal subunit concentration due to capsid excluded volume.}\label{sec:excluded_volume}
 Figs.~\ref{fig:yield_vs_concentration_dodecahedron} and~\ref{fig:yield_vs_concentration_icosahedron} show the long (but finite) time yields as a function of total subunit concentration ($\rhoT$) for the dodecahedron and icosahedron models, along with representative simulation snapshots. Results are shown for several values of the partition coefficient for a relatively low binding affinity, $\ebind=6$. While we observe almost no assembly in the absence of  LLPS (see also Supplementary Movie S1), except for low yields at high $\rhoT$, LLPS leads to high yields over a wide range of concentrations. However, there is an optimal value of $\rhoT$ beyond which yields decrease. For small partition coefficients, the optimal $\rhoT$ decreases with increasing partition coefficient, but then seems to saturate; e.g. at $\rhoT\approx 3\times10^{-4}$ for the dodecahedron model. Note that the saturation value depends on $\ebind$ and $\Vr$. The following analysis shows that these trends result from a combination of thermodynamic and kinetic effects.

First, we compare the simulation results against the equilibrium ideal solution theory from  Ref.~\cite{Hagan2023} (see SI Section S3 of this article). The dotted line in Fig.~\ref{fig:yield_vs_concentration_dodecahedron} shows the approximate equilibrium ideal solution result for the highest partition coefficient ($\Econd=7$), and Fig. S1 shows results for different values of $\Econd$. While qualitatively correct for low $\rhoT$, the ideal solution theory fails to capture the decrease in yields for large $\rhoT$. The ideal solution rate equation model also predicts high yields in this regime \cite{Hagan2023}. 

While previous simulation results in bulk solution exhibit such decreases in yield due to monomer starvation or malformed structures, we observe that almost all subunits are either monomers or part of assembled capsids for the entire range of $\rhoT$ and $\Kc$, with vanishingly few intermediates or malformed structures (see Fig. S7). Instead, we can explain the decrease in yield as a \textit{thermodynamic} consequence of subunit excluded volume. For $\rhoT \gtrapprox5\times 10^{-4}$, the concentration of assembled capsids within the condensate is roughly constant (Fig. S8), consistent with snapshots which show capsids densely packed within condensates and no capsids in the background (see Figs.~\ref{fig:yield_vs_concentration_dodecahedron},~\ref{fig:yield_vs_concentration_icosahedron}). Capsids are near the close-packing limit, and hence the condensate cannot accommodate higher capsid concentrations as the total subunit concentration increases. These images are strikingly reminiscent of recent observations of rotavirus capsids closely packed in viroplasms (see \ref{fig:sim_vs_expt}). 

Based on this observation, we developed a theory to predict the equilibrium capsid concentration, which accounts for subunits’ excluded volume by approximating them as hard spheres (see SI Section S3). To briefly summarize the theory, we make the two-state approximation that subunits exist either as monomers or as part of assembled capsids. Equilibrium implies that the chemical potentials of monomers and capsids within each phase are equal, and that the chemical potentials of each species in the condensate and background phases are equal. Assuming that we can model monomer and capsid excluded volume by treating each of these species as effective hard spheres (whose diameters we estimate in SI Section S4), we estimate the chemical potentials using a standard hard sphere equation of state~\cite{Mansoori1971}, and solve the resulting equations self-consistently to obtain the equilibrium concentrations of monomers and capsids in the condensate and background. The hard sphere theory results are shown (dashed line) for $\Econd=7$ in Figs.~\ref{fig:yield_vs_concentration_dodecahedron},~\ref{fig:yield_vs_concentration_icosahedron} and several values of $\Econd$ in  Fig. S1. 
The theory correctly captures the non-monotonic dependence of yield on $\rhoT$, anticipates that the capsid concentration within the condensate saturates as $\rhoT$ increases (Fig. S8), and predicts the
 optimal value of $\rhoT$ to within a factor of 2 ($\approx 2\times 10^{-4}$ in theory vs $\approx 4\times 10^{-4}$ in simulations). For low $\rhoT$ the theoretical yields are above the simulation results, which can be attributed to the fact that the simulation results are finite-time and thus smaller than equilibrium (assembly reactions approach equilibrium asymptotically slowly due to increasing nucleation barriers as subunits are depleted, see Fig. S4 and SI Section S5) \cite{Hagan2010, Hagan2014}) and that we only roughly estimate the subunit interaction free energy within capsids (see SI Section S4). 
The theory is less accurate, though still qualitatively good, at high $\rhoT$ where capsid concentrations approach the close-packing limit (Fig. S8). 
We do not expect quantitative agreement in this regime because the Carnahan-Starling approximation significantly deviates from real hard-sphere behavior for $\eta \gtrsim 0.5$~\cite{Carnahan1970,Mansoori1971}. Note that the slight increase in the theoretical yield at $\rhoT \gtrsim 8\times 10^{-4}$ arises due to assembly in the background, which we do not observe in simulations due to nucleation barriers.

Notably, there is a broad range of $\rhoT$ over which the total concentration of assembled capsids $\rhon \approx \rhonc \Vr$  is roughly constant, although at high enough concentrations (above the bulk critical subunit concentration ($\rhoStar$) capsids will also assemble in the background. Thus, condensates may provide a mechanism to control the total number of assembled capsids, since the number of capsids is bounded by the total condensate volume within this regime. 
Interestingly, optimal yields occur for capsid concentrations slightly below the close-packing limit, $\rhonc/\rhocp\approx 0.6$. This observation reflects thermodynamic costs (low translational entropy of capsids) and kinetic factors (slow subunit diffusion and hence assembly rates) at the close-packing limit.

\subsubsection{There is an optimal condensate volume fraction due to capsid excluded volume.}

The simulation yields also depend nonmonotonically on the condensate volume fraction $\Vr$ (Fig.~\ref{fig:yield_vs_vr}), with an optimal value of $\Vr^*$ that decreases with $\Econd$, and a sharp decrease in yields and finite-time yield for $\Vr\lesssim 5\times10^{-3}$. This behavior is nearly quantitatively captured by the hard sphere theory for high $\Econd$, but not by ideal solution theory, which predicts that yields increase monotonically with decreasing $\Vr$ (see Fig. S9 for theoretical predictions at different $\Econd$). The origin of this behavior is evident from simulation snapshots, which show that the condensate becomes highly packed with assembled capsids for high $\Econd$ and low $\Vr$.

In fact, the yields in this regime and the point corresponding to the steep decline can be qualitatively estimated from a simple sphere packing argument. Although the capsids have dodecahedral symmetry, they have a rounded shape and thus we treat them as effective spheres rather than faceted polyhedra in the following argument. Neglecting free monomers under the assumption of high yield, the maximum number of subunits that can form capsids within a condensate is $N_\text{max} \approx \ncap\Vc \rhocp$, where $\rhocp$ is the capsid close-packing concentration:
\begin{equation}
\rhocp=\pi/(3\sqrt{2})/(4\pi(\sigcap/2)^3/3)\approx 1.28\times10^{-2}
\end{equation}
where the capsid excluded volume size $\sigcap=5.1$ is estimated in SI Section S4A. 
Thus, below a threshold volume fraction $\Vr^*$  the total number of subunits exceeds the maximum number in the condensate $N=V\rhoT>N_\text{max}$ and the yield declines (assuming assembly only occurs in the condensate). The threshold is
$\Vr^*\approx 1/\left(\frac{\ncap\rhocp}{\rhoT}-1\right)$,
and the corresponding yield for $\Vr\leq\Vr^*$ is
\begin{equation}
    \fcp = \frac{(\Vr^*)^{-1}+1}{\Vr^{-1}+1}.\label{eq:yield_fcp}
\end{equation}
We plot $\fcp$ in Fig.~\ref{fig:yield_vs_vr} and indicate the value of $\Vr^*(\approx 2.61\times 10^{-3})$. The theory qualitatively matches the simulation results, although the decline occurs at slightly larger $\Vr$ in both finite-time simulation results and the equilibrium hard sphere theory. This agreement suggests that the simple packing argument explains the decline in yield at low $\Vr$, but that assembly driving forces are not strong enough to reach complete close-packing. Indeed, Fig. S8 shows that capsid concentrations appear to saturate below the close-packing density, reaching $\rhonc/\rhocp\approx 0.8$ for $\Econd=7$, $\rhoT=10^{-3}$.  

\subsubsection{There is an optimal binding affinity due to nucleation barriers and malformed structures.}
Fig.~\ref{fig:yield_vs_ess}A shows the long-time yields as a function of the subunit binding affinity parameter $\ebind$ for the dodecahedron model, along with representative simulation snapshots. We observe that LLPS significantly enhances yields, particularly at low $\ebind$ for which we observe no assembly in the absence of phase separation. Notably, we even observe higher yields for binding affinities that lead to strong assembly without phase separation ($\ebind=7$; see Supplementary Movie S2). However, for $\ebind>7$ there is a sharp decrease in yields for all partition coefficients. In contrast to the case of high subunit concentration, this decline is not captured by the hard sphere theory (shown for $\Econd=7$), indicating that it  does not arise due to excluded volume. Instead, the discrepancies arise due to kinetic effects. For low binding affinities, nucleation barriers prevent reaching equilibrium within achievable simulation timescales. Since the nucleation barrier increases with decreasing binding affinity, the simulation results deviate further from equilibrium as $\ebind$ decreases. However, as we discuss in section \ref{sec:rate}, LLPS has a significant effect on nucleation timescales. 

In contrast, as the binding affinity increases beyond $\ebind=7$, nucleation barriers are relatively small, but the prevalence of malformed off-pathway structures steadily increases (Fig.~\ref{fig:yield_vs_ess}B). The appearance of malformed structures at strong binding affinity values is consistent with previous computational and experimental results from assembly in bulk solution \cite{Hagan2006,Hagan2011,Hagan2014,Zlotnick1999,Zlotnick2000,Sorger1986,Parent2007a,Stray2005,Kondylis2018}. Indeed, for $\ebind\geq7$, we observe malformed structures (as well as assembled capsids) in both the background and the condensate (see Supplementary Movies S6, S7). However, Fig.~\ref{fig:yield_vs_ess} exhibits several notable features that highlight the interplay between malformed structures and LLPS-coupled assembly. First, the fraction of subunits in off-target structures is insensitive to $\Econd$. This result is unexpected given the high local concentrations of subunits in the condensate, since high concentrations also favor the formation of malformed structures in bulk solution. Second, although the enhancement of yields is much greater for smaller binding affinities, LLPS enhances yields even within the malformed assembly regime. We attribute this enhancement to the ability of LLPS to prevent the monomer starvation trap, as discussed in Section~\ref{sec:rate}. Third, the structures of malformed assemblies in the condensate differ from those in bulk solution. In bulk, malformed structures usually arise because subunits bind with incorrect geometries and are not significantly larger than well-formed capsids. For example, we commonly observe assemblies with 12 subunits but only 26 bonds. These ``danglers'' \cite{Trubiano2024a} (Fig. S10) form when the twelfth subunit binds in the wrong orientation; the unbinding is slow at high $\ebind$. In the condensate we observe fewer danglers (Fig. S10), but frequently see partially assembled capsids bound together to form large aggregates (snapshot (iii) in Fig.~\ref{fig:yield_vs_ess} and Supplementary Movie S7).  
This difference is reflected in a much larger average cluster size for $\ebind\geq7.5$ for LLPS-coupled assembly compared to $\Econd=0$ (Fig. S11). We attribute these aggregates to the high local concentration of intermediates within the condensate under these conditions; binding of large intermediates to each other frequently leads to malformed structures \cite{Whitelam2009}.

\subsection{LLPS enhances assembly rates}\label{sec:rate}

\begin{figure*}[ht]
    \centering
    \includegraphics[width=1.0\linewidth]{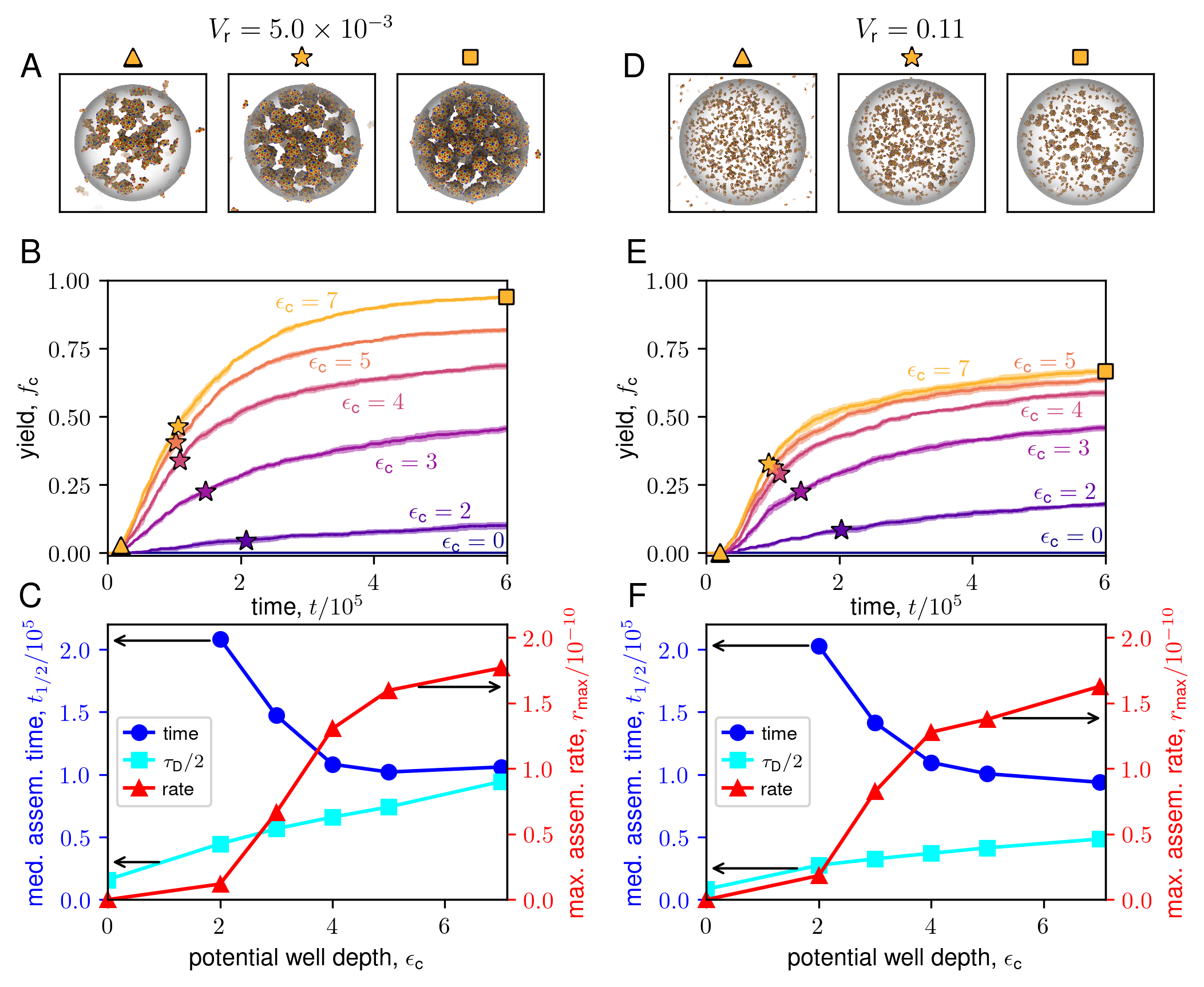}
    \caption{LLPS-coupled assembly kinetics.  The left column (panels A-C) shows results for $\Vr=5.0\times10^{-3}$, and the right column (panels D-F) shows results for $\Vr=0.11$. (A,D) Snapshots from assembly trajectories with $\Econd=7$, corresponding to times $t/10^5=0.2,1$, and 6, indicated by triangle, star, and square symbols on the plots in (B, E). See Supplementary Movies S4, S5 for the corresponding trajectories. (B,E) Yield versus time for different values of $\Econd$. Stars indicate the median assembly times at all $\Econd$. Parameters for both values of $\Vr$ are: $\ebind=6$, $\rhoT=4\times10^{-4}$. (C,F) Median assembly time, $\tHalf$ (blue, circles), and maximum assembly rate, $\rMax$ (red, triangles), versus $\Econd$. The solid cyan line with squares indicates half the diffusion time, $\tDiff/2$ (Eq.~\eqref{eq:tDiff}). Arrows indicate the $y$ axis to which the lines correspond.
    }
    \label{fig:yield_vs_time}
\end{figure*}

\begin{figure}
    \centering
    \includegraphics[width=1.0\linewidth]{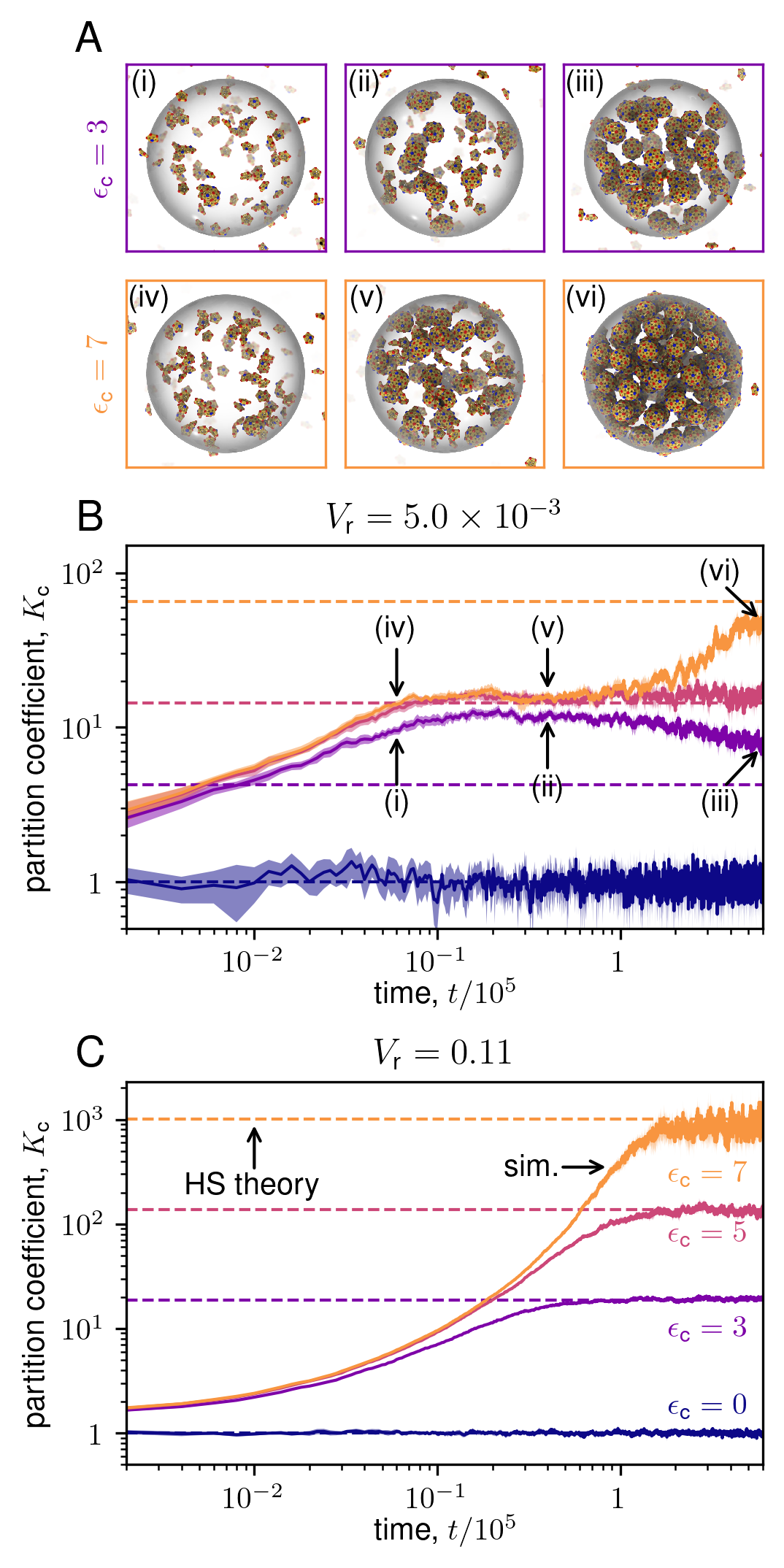}
    \caption{Partitioning of subunits into condensate during assembly. (A) Snapshots from assembly trajectories for $\Vr=5.0\times10^{-3}$ and $\Econd=3$ (top) and 7 (bottom). See Supplementary Movies S3, S4 for segments of the corresponding trajectories. (B) Partition coefficient $\KcActual(t)=\rhoC(t)/\rhoB(t)$ measured in simulations as a function of time for indicated values of $\Econd$ at $\Vr=5.0\times10^{-3}$. Solid lines are $\KcActual(t)$, and dashed lines are the predictions of HS theory. For $\Econd=3$, $\KcActual$ initially rises well above the value of $\Kc$ appropriate for equilibrium between subunits and capsids, instead approaching the equilibrium value of $\Kc$ for no assembly (see Fig.~\ref{fig:Kc_vs_eps_c}). (C) $\KcActual(t)$ for indicated values of $\Econd$ at $\Vr=0.11$.} 
    \label{fig:Kc_vs_time}
\end{figure}

In addition to enhancing long-time yields, LLPS also significantly increases assembly rates and thus the range of parameters leading to productive assembly at relevant timescales. Fig.~\ref{fig:yield_vs_time}A,C show the yield as a function of time for different values of $\Econd$ for moderate and large condensate volume fractions respectively ($\Vr=5.0\times10^{-3}$ and 0.11), for fixed  $\ebind=6$ and $\rhoT=4\times10^{-4}$. Fig.~\ref{fig:Kc_vs_time}A shows representative simulation snapshots at various times for $\Econd=7, 3$. 
While there is no assembly without LLPS ($\Econd=0$) at these conditions, we observe rapid and productive assembly with LLPS. Rates increase dramatically with $\Econd$ for weak partitioning, but begin to level off for $\Econd > 3$. More notably, the median assembly timescales $\tHalf$ saturate at a minimum value of $\tHalfMin \approx 10^5$. 

To further quantify these trends, we plot the median assembly timescales and maximum assembly rates $\rMax$ as a function of  $\Econd$ for moderate and large condensate volume in Fig.~\ref{fig:yield_vs_time}B,D. We estimate the maximum assembly rate by convolving the assembly kinetics with the first derivative of a Gaussian (see Fig. S12). We consider the maximum assembly rate as an estimate of the initial nucleation rate because it eliminates the lag time required for a nucleus to grow into a complete capsid \cite{Hagan2010}.

The observed acceleration of assembly rates with increasing $\Econd$ and decreasing $\Vr$ is consistent with the prediction from the rate equation theory  \cite{Hagan2023}. However, the theory fails to capture the saturation of rates at high  partitioning strength  (see SI section S6 and Fig. S5). 

The discrepancy between the ideal solution rate equation and the computational results arises because the interplay between assembly, capsid excluded volume, and subunit diffusion rates leads to a complex dynamics of subunit concentrations within the condensate for high $\Econd$ and low $\Vr$. Fig.~\ref{fig:Kc_vs_time} shows the partition coefficient $\KcActual(t)=\rhoC(t)/\rhoB(t)$ measured in simulations as a function of time for some of the parameter sets in Fig.~\ref{fig:yield_vs_time}. For large condensates ($\Vr=0.11$, Fig.~\ref{fig:Kc_vs_time}C) subunits steadily partition into the condensate, reaching the equilibrium value of the partition coefficient by $t=2\times 10^5$. In contrast, for $\Vr=0.005$ (Fig.~\ref{fig:Kc_vs_time}B) and high $\Econd \gtrsim 4$, subunit concentrations stall at $\KcActual \approx 10$; and then only gradually increase at later times. This is consistent with the observation that median assembly timescales saturate for  $\Econd \gtrsim 4$.
Intriguingly though, $\KcActual$ overshoots the equilibrium value at early times for $\Econd=3$.

We can understand the results at low $\Vr$ as follows. Simulation trajectories show that assembly occurs extremely rapidly for low $\Vr$ and high partitioning $\Econd\gtrsim 7$ (see snapshot (v) in Fig.~\ref{fig:Kc_vs_time}A), depleting subunits fast enough that the subunit flux from diffusion into the condensate ($\approx 4 \pi R D \rhoB$ with $D=\kt/\gamma$ the subunit diffusion constant) cannot maintain the equilibrium partition coefficient (see Fig.~\ref{fig:Kc_vs_time}B). Thus, the condensate reaches a quasi-steady-state where assembly rates are balanced by the subunit flux into the condensate, resembling previous models of assembly in the presence of monomer influx (\cite{Hagan2011, Castelnovo2014}).
 Then, as the condensate fills with capsids, the high excluded volume further slows subunit entry, leading to an extremely long timescale for subunit densities to reach equilibrium. In contrast, for weaker partitioning $\Econd \lesssim 3$, nucleation in the condensate is sufficiently slow that subunit partitioning is unimpeded by capsid excluded volume at early times. Thus, subunit concentrations approach the equilibrium partition coefficient in the \textit{absence} of capsid excluded volume (see Fig.~\ref{fig:Kc_vs_eps_c}), leading to the apparent overshoot. Then, as assembly proceeds, capsid excluded volume increases, and $\rhoC$ decreases toward the equilibrium value. 

These considerations show that for sufficiently strong partitioning, reaction rates will eventually be limited by the diffusive flux of subunits into the condensate. We can estimate the characteristic timescale for subunit diffusion into the condensate as (see SI section S6)
\begin{align}
\tDiff \approx \frac{\Vc/(\Vr+1/\Kc) + \fc \VT}{4\pi \Rc D}
\label{eq:tDiff}
\end{align}
where the first and second terms in the numerator respectively account for diffusion of free subunits and those that eventually form capsids. As shown in Fig.~\ref{fig:yield_vs_time}, $\tDiff/2$ (the cyan curve, with the factor of $\frac{1}{2}$ because it is the median time) closely matches the minimum median assembly timescale for strong partitioning. Thus, subunit diffusion creates a speed limit on LLPS-facilitated assembly and thus a lower bound on assembly timescales.

\textit{\textbf{Implications for rate equation theory.}}
The above considerations show that the ideal solution rate equation theory needs to be extended to account for excluded volume effects as capsids accumulate in the condensate, which lead to reduced partition coefficients (as described in section~\ref{sec:condensate_model}) and slower subunit diffusion with correspondingly slower reaction rates. 

In addition, we previously proposed a simple scaling estimate that LLPS accelerates nucleation rates and decreases assembly timescales by a factor (see Ref. \cite{Hagan2023} and SI section S6):
\begin{align}
\sNuc \approx  & \Vr / \left(\Vr + 1/\Kd\right)^{\nNuc} \quad \mbox{for } \Vr \ll 1.
\label{eq:snuc}
\end{align}
This estimate assumed an ideal solution, a critical nucleus size $\nNuc$ that is independent of conditions, and that subunit diffusion is fast in comparison to assembly timescales so that the partition coefficient $\Kc$ maintains its equilibrium value. The simulations show that all these assumptions need to be relaxed at strong partitioning. We show in SI section S6 and Fig. S5 that this scaling estimate can  be made to qualitatively match simulation results by using the measured values of the partition coefficient $\KcActual$ in $\sNuc$, and extending the expression for the median assembly timescale to include the diffusion timescale:
\begin{align}
\tHalf \approx \tHalf^0/\sNuc + \tDiff/2
\label{eq:tHalf}
\end{align}
with $\tHalf^0$ the timescale in the absence of LLPS.  Further effects that need to be accounted for include the fact that the critical nucleus size can decrease as local subunit concentrations increase and the above-mentioned decrease in reaction rates due to excluded volume.

\textit{\textbf{LLPS enables high rates by creating a buffer of free subunits.}}
By spatially localizing assembly within the condensate, LLPS enables extremely high assembly rates while limiting the overall rate of consuming subunits for $\Vr \lesssim 0.1$. Under these conditions, the background solution acts as a  ``buffer'' that steadily supplies free subunits to the condensate. This effect can avoid the monomer starvation kinetic trap, thus greatly increasing the possible rates and yields of assembly.  

\begin{figure}[hbtp]
    \centering
    \includegraphics[width=1.0\linewidth]{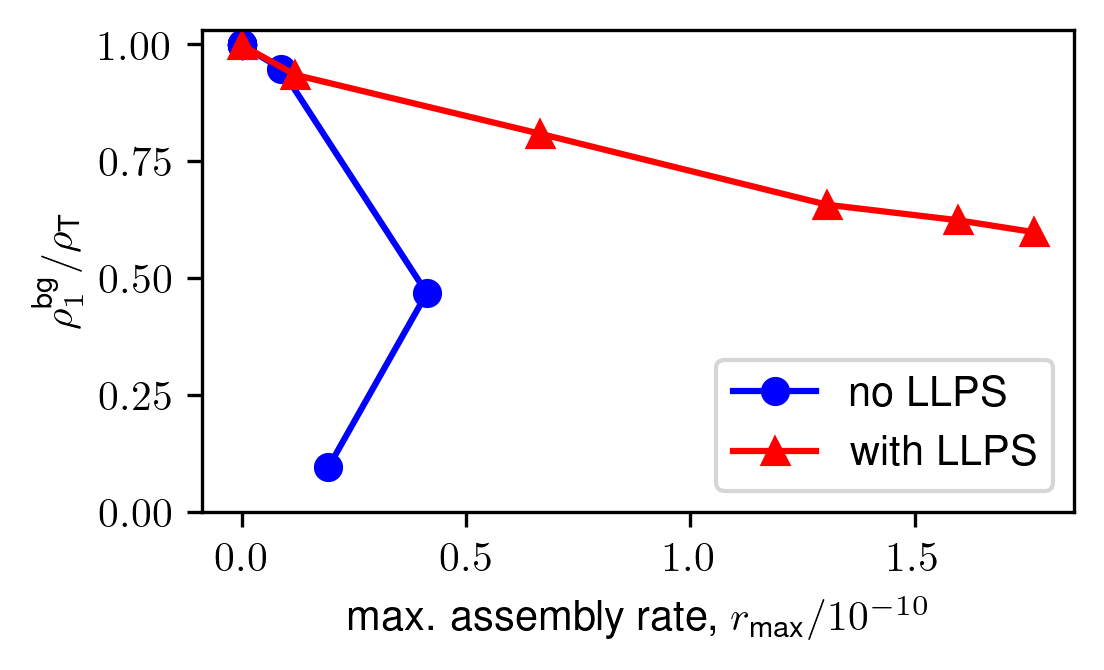}
    \caption{The bulk solution acts as a buffer of free subunits with LLPS. The plot shows the background concentration $\rhoB$ (normalized by the total concentration $\rhoT$) as a function of the maximum assembly rate both with (red, triangles) and without (blue, circles) LLPS. Each data point for ``no LLPS'' corresponds to a different value of $\ebind$, which ranges from $4 \le \ebind \le 8$, and $\Econd=0$. Each data point for ``with LLPS'' corresponds to a different $\Econd$, with $\ebind=6$.}
    \label{fig:max_rate}
\end{figure}

We quantify this mechanism by plotting the concentration $\rhoB$ of free subunits in the background as a function of the maximum assembly rate, $\rMax$, for different values of $\Econd$ over a range of $\ebind$ (Fig.~\ref{fig:max_rate}). Each data point in the plot corresponds to a different value of $\ebind$ for the indicated $\Econd$ (shown in different colors). We see that for no LLPS ($\Econd=0$), $\rhoB$ decreases rapidly with increasing $\rMax$ (which is achieved by increasing $\ebind$) because rapid nucleation throughout the entire simulation box depletes free subunits. In contrast, with LLPS ($\Kc>1$), $\rhoB$ remains fairly close to $\rhoT$ for assembly rates that are an order of magnitude higher, thus acting as a buffer. However, even with LLPS there is eventually a binding affinity above which the maximum rate decreases, due to assembly in the background solution, formation of malformed structures, or overly rapid assembly within the condensate.

\section{Discussion}\label{sec:discussion}

While these results are qualitatively consistent with previous rate equation models \cite{Hagan2023,Bartolucci2024}, the particle-based simulations exhibit important differences due to excluded volume effects and aberrant off-pathway structures that are not accounted for in ideal solution theory and simplified rate equations. In particular, there is an optimal total subunit concentration $\rhoT$ and condensate volume fraction $\Vr$. For higher total subunit concentrations or lower condensate volumes, yields are limited by capsid excluded volume in the condensate. We developed an equilibrium theory for capsid assembly that includes these excluded volume effects, which agrees well with the simulation results. For higher-than-optimal binding affinity values, yields are suppressed by malformed off-pathway structures, independent of the subunit partition coefficient $\Kc$. However, malformed structures form large aggregates within the condensate, while they tend to be on the order of the capsid size in bulk. In addition, we find that assembly rates are limited at high partition coefficients by the rate of subunit diffusion into the condensates. 
Rates are then further reduced by slow subunit diffusion once capsid excluded volume in the condensate becomes large. The latter effect is consistent with previous theoretical models \cite{Ponisch2021,Schmit2024}. We show how the rate equation and associated scaling estimate for LLPS-facilitated assembly rates  and timescales \cite{Hagan2023} can be extended to include these effects.

\textit{Diffusion sets a speed limit for LLPS-facilitated assembly.} For sufficiently strong partition coefficients, leading to high local subunit concentrations within the condensate, assembly rates can become limited by the diffusive flux of subunits into the condensate. The associated diffusive timescale sets a  lower bound on the timescale for LLPS-facilitated assembly.

\textbf{\textit{Testing in experiments.}}
Our predictions on the dependence of yield and assembly rates on subunit concentrations and binding affinity values could also be tested in both biological and synthetic experiments in which self-assembly occurs within phase-separated condensates. In addition to viruses --- the system that most closely motivates this work --- our predictions could also be tested other cases described in the introduction (e.g. clathrin cages, actin filaments, and assemblies within neuronal synapses \cite{Day2021, Zeng2018a, McDonald2020, Emperador-Melero2021, Lipinski2022, McCall2018}).  However, these predictions may be more readily tested in \textit{in vitro} experiments which allow greater control over parameter values and condensate sizes, such as biomolecular assemblies \cite{Bergeron-Sandoval2016,DelSignore2021,Guseva2020,Banani2017}, or recently developed DNA origami subunits that form capsids and tubules \cite{Videbaek2022,Videbaek2024,Videbaek2025,Wei2024,Sigl2021}.    

The computational prediction that capsids fill phase-separated compartments to near close-packing densities under optimal assembly conditions are evocative of recent observations from cells infected with rotavirus \cite{Papa2021,Geiger2021}. It would be of interest to investigate whether similar capsid arrays are observed in other viruses, what conditions give rise to such arrays, and more generally how capsid packing densities depend on parameter values.

\textbf{\textit{Implications for biology and synthetic assembly systems.}}
The ability of LLPS to increase assembly rates could be crucial for viruses to assemble before detection by host immune systems, and thus may be one reason why many viral systems form phase-separated compartments during their lifecycles. Further, while viruses have limited control over conditions within the host cell, they can control the partition coefficient and condensate volume. Faster assembly rates and enhanced control may also explain why many other biological assembly processes occur within condensates.  In addition, the computational results show that there is a broad range of concentrations over which capsid concentrations within the condensate are nearly constant. Thus, by controlling the condensate volume, viruses can control the total number of assembled capsids. This might also provide a mechanism to control the timing of assembly --- additional assembly would occur whenever capsids are released from the condensate. 

Importantly, these same considerations suggest that LLPS could provide a means to enable rapid, robust assembly in human-engineered systems. In addition, our study identifies a route for efficient bottom-up assembly of highly monodisperse arrays of assemblies, which could have important optoelectronic applications. In contrast to previous approaches that can only assemble macroscopic crystals of capsids \cite{Asor2017,Uchida2018,McCoy2018}, condensate-coupled assembly enables precisely controlling the size of the crystalline arrays.

\textbf{\textit{Outlook.}}
The excluded volume effects identified in this work could be, at least approximately, accounted for in rate equation models following the approach that we have used for the equilibrium theory. Moreover, the theory can be applied to other assembly reactions occurring at high concentrations, such as the assembly of the mature HIV capsid \cite{Ganser-Pornillos2004,Ganser-Pornillos2008,Yu2013,Nermut2003,Adamson2004,Briggs2006,Chen2011,Krishna2010,Baumgaertel2012,Balasubramaniam2011,Zhao2013,Grime2012,Gupta2023}. Although our computational model accounts for this and other effects that were neglected in previous rate equation models, we have only implicitly modeled the condensate and assumed that the condensate phase coexistence is independent of subunit concentration. A natural next step is to explicitly model the constituents that form the condensate. This will allow studying the effect of subunits on condensate phase separation as well as additional effects that may arise from the excluded volume of condensate components.

\begin{acknowledgments}
    This work was supported by the NSF through DMR 2309635 and the Brandeis Center for Bioinspired Soft Materials, an NSF MRSEC (DMR-2011846). Computing resources were provided by the National Energy Research Scientific Computing Center (NERSC), a Department of Energy Office of Science User Facility (award BES-ERCAP0026774); the NSF ACCESS allocation TG-MCB090163; and the Brandeis HPCC which is partially supported by the NSF through DMR-MRSEC 2011846 and OAC-1920147.
\end{acknowledgments}

\section*{Conflicts of Interest}
The authors declare no conflicts of interest. 

\section*{Data Availability Statement}
Our simulation and analysis scripts are available on Github: https://github.com/Layne28/implicit-condensate.
Additionally, post-processed trajectory data, Mathematica notebooks for theoretical calculations, and figure-generating scripts are hosted on the Open Science Framework OSFHome (https://osf.io/hq2y8/). 

\appendix

\section{Partition coefficient versus $\Econd$}

\begin{figure}[hbtp!]
    \centering
    \includegraphics[width=1.0\linewidth]{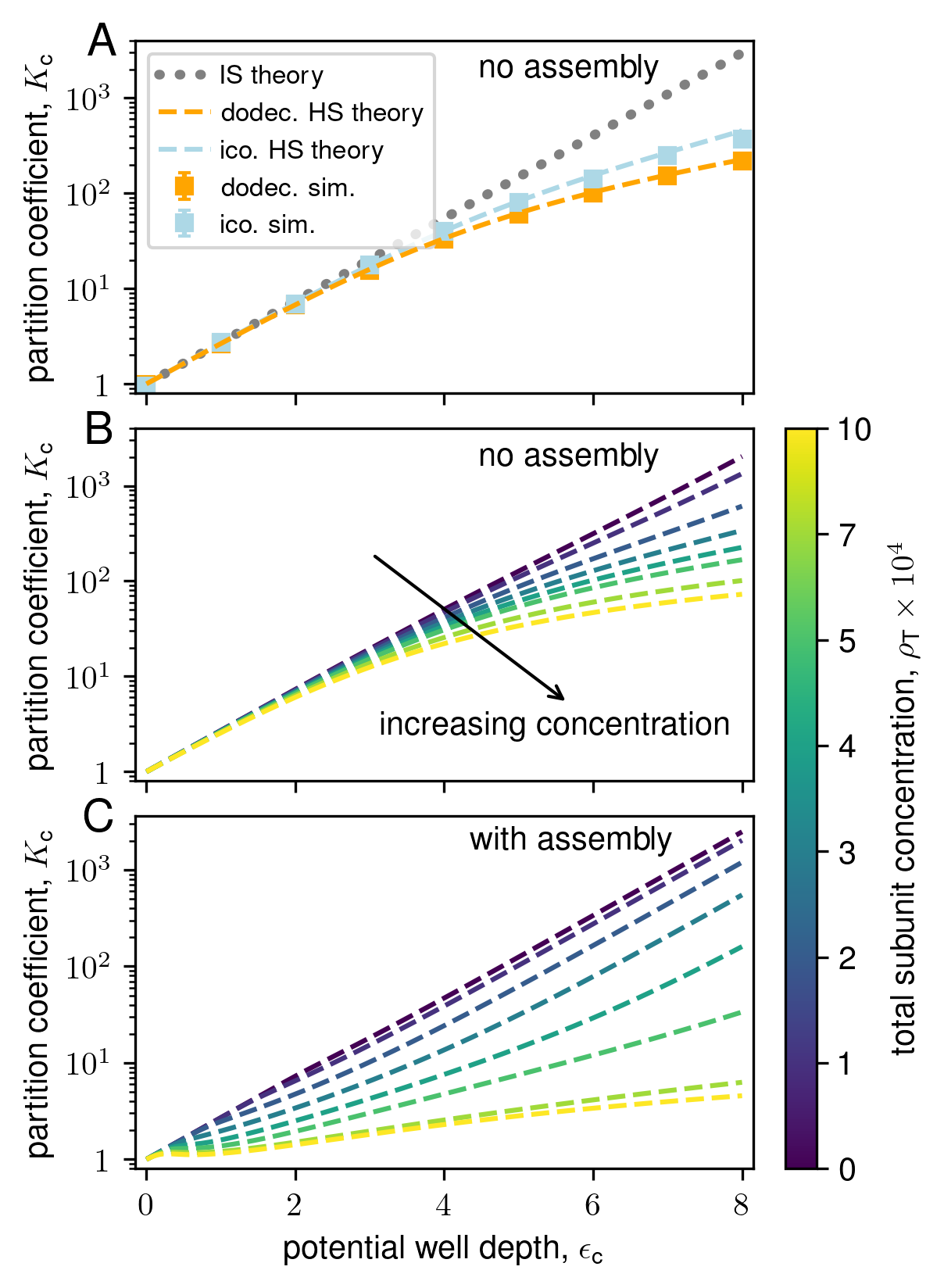}
    \caption{Effects of excluded volume on the partition coefficient, with and without assembly. (A) Partition coefficient $\Kc$ versus condensate potential well depth $\Econd$ with no assembly (i.e. no subunit-subunit attractions, $\ebind=0$). The gray dotted line represents ideal solution (IS) theory. Orange and light blue represent results for the dodecahedron and icosahedron models, respectively, at concentrations $\rhoT=4\times10^{-4}$ (dodecahedron) and $\rhoT=7.5\times 10^{-5}$ (icosahedron). The dashed lines represent the hard sphere (HS) theory using the Carnahan-Starling equation of state (see Section S2), and the square markers represent results from simulations with $\ebind=0$ (error bars are smaller than the size of the plotted points). In both cases, ideal solution theory breaks down when packing fractions in the condensate exceed $\eta \gtrsim 10\%$ (Fig. S6). (B) Hard sphere theory results for the dodecahedron model without assembly over a range of concentrations. At the highest well depth and subunit concentration, excluded volume causes $\Kc$ to decrease by over an order of magnitude compared to ideal solution theory. (C) Hard sphere theory results for the dodecahedron model with assembly over a range of concentrations. At high concentrations, excluded volume causes $\Kc$ to decrease by orders of magnitude compared to ideal solution theory predictions.
    }
    \label{fig:Kc_vs_eps_c}
\end{figure}

Fig.~\ref{fig:Kc_vs_eps_c} shows the partition coefficient $\Kc$ as a function of potential well depth $\Econd$ for both capsid models at a single concentration with no assembly, and for the dodecahedron model with and without assembly over a range of concentrations. 

\bibliography{references}

\end{document}

\title{Supplementary Information: Computer simulations show that liquid-liquid phase separation enhances self-assembly}

\author{Layne B. Frechette}
\thanks{These two authors contributed equally.}
\affiliation{
 Martin Fisher School of Physics, Brandeis University, Waltham, Massachusetts 02453, USA
}

\author{Naren Sundararajan}
\thanks{These two authors contributed equally.}
\affiliation{
 Martin Fisher School of Physics, Brandeis University, Waltham, Massachusetts 02453, USA
}

\author{Fernando Caballero}
\affiliation{
 Martin Fisher School of Physics, Brandeis University, Waltham, Massachusetts 02453, USA
}
\author{Anthony Trubiano}
\affiliation{
 Martin Fisher School of Physics, Brandeis University, Waltham, Massachusetts 02453, USA
}
\author{Michael F. Hagan}
\email{hagan@brandeis.edu}
\affiliation{
 Martin Fisher School of Physics, Brandeis University, Waltham, Massachusetts 02453, USA
}

\date{\today}

\maketitle
\onecolumngrid
\tableofcontents

\section{Model description}\label{supp_sec:model_description}

\subsection{Dodecahedral capsid model}
In the dodecahedron model, each subunit consists of seven pseudoatoms of three different types. Five attractor pseudoatoms (`A'), which facilitate subunit-subunit binding, are located at the vertices of of a regular pentagon with circumradius $l_0$, which we take as the unit of length for this model. `A' pseudoatoms on different subunits interact via a Morse potential:
\begin{equation}
    u_{\text{AA}}(r) = \begin{cases}
    \ebind\left[e^{-2\alpha(r-r_0)} - 2e^{-\alpha(r-r_0)}\right] -\epsilon_0, &r\leq 2l_0 \\
    0, &r>2l_0
    \end{cases} \label{eq:uaa}
\end{equation}
where $r$ is the distance between `A' pseudoatoms, $r_0=l_0/5$ is the distance at which the potential energy is minimized, $\alpha=2.5/r_0$ sets the width of the potential well, and $\epsilon_0=\ebind\left[e^{-2\alpha(2l_0-r_0)}-2e^{-\alpha(2l_0-r_0)}\right]$ is chosen so that the potential goes to zero continuously at $r=2l_0$. The parameter $\ebind$ sets the depth of the potential well and hence controls the subunit-subunit binding affinity. Additionally, each subunit contains one top pseudoatom (`T') and one bottom pseudoatom (`B') located at $z=\pm l_0/2$ relative to the centroid of the pentagon, which lies in the $xy$ plane in the body-fixed frame. `T' pseudoatoms interact via a purely repulsive potential:
\begin{equation}
    u_{\text{TT}}(r) = \mathcal{L}(r;\sigma_{\text{TT}},\epstt), \label{eq:utt}
\end{equation}
where: 
\begin{equation}
    \mathcal{L}(r;\sigma,\epsilon) = \begin{cases}
     4\epsilon\left[\left(\frac{\sigma}{r}\right)^{12}-\left(\frac{\sigma}{r}\right)^6\right], & r\leq\sigma \\
     0, & r>\sigma.
     \end{cases}
\end{equation}
We choose $\epstt=1$ and $\sigtt=2.1l_0$, which encourages a subunit-subunit binding angle consistent with a dodecahedron. `B' pseudoatoms interact with `T' pseudoatoms via a similarly repulsive potential:
\begin{equation}
    u_{\text{TB}}(r) = \mathcal{L}(r;\sigtb,\epstb), \label{eq:utb}
\end{equation}
where $\epstb=\epstt$ and $\sigtb=1.8l_0$. The `T'-`B' interactions help prevent ``upside down'' assembly, by discouraging subunit-subunit binding that results in `T' and `B' pseudoatoms on different subunits being adjacent. Subunits and their interactions are depicted in Fig. 2A,B, and a dodecahedral capsid resulting from the assembly of twelve subunits is shown in Fig. 2C. SI Section~\ref{supp_sec:param_estimate}B provides an estimate of the subunit-subunit binding affinity and capsid free energy as a function of $\ebind$.

\subsection{Icosahedral capsid model}
The icosahedron model was previously developed to study the assembly of DNA origami capsids \cite{Sigl2021, Wei2024}. The subunit shape and the interactions between the constituent pseudoatoms are designed such that the minimum energy configuration of a collection of twenty subunits is an icosahedral shell (see Fig. 2F). Subunits consist of 45 purely repulsive excluder pseudoatoms and six attractor pseudoatoms. As in the dodecahedron model, subunit-subunit binding affinities are governed by the parameter $\ebind$. However, unlike the dodecahedron model, not all attractor atoms interact with each other equally. Instead, each attractor pseudoatom within a subunit has a distinct type (represented as different colors in Fig. 2D,E). Pairs of complementary attractor atoms attract each other, while non-complementary attractor atoms interact only via excluded volume. This prevents subunits from binding ``upside down'' (thus it has a similar role to the repulsive `T'-`B' interactions in the dodecahedron model). Complementary attractor pairs are listed in Table \ref{table:interactions}.

\begin{table}[ht]
\caption{Complementary attractors \\(1=complementary, 0=non-complementary)} 
\centering
\begin{tabular}{c c c c c c c} 
\hline\hline 
 & blue & red & gray & orange & yellow & tan \\ [0.5ex] 
blue & 0 & 1 & 0 & 1 & 0 & 1\\ 
red & 1 & 0 & 1 & 0 & 1 & 0\\
gray & 0 & 1 & 0 & 1 & 0 & 1\\
orange & 1 & 0 & 1 & 0 & 1 & 0\\
yellow & 0 & 1 & 0 & 1 & 0 & 1\\
tan & 1 & 0 & 1 & 0 & 1 & 0\\
\hline 
\end{tabular}
\label{table:interactions} 
\end{table}

Excluder pseudoatoms and non-complementary attractor pseudoatoms interact via a Weeks-Chandler-Anderson (WCA) potential \cite{Weeks1971}:
\begin{equation}
    \uwca(r) = \begin{cases} 4 \ebind \left[\left(\frac{\sigma}{r}\right)^{12}-\left(\frac{\sigma}{r}\right)^6+\frac{1}{4}\right], & r < 2^{1/6}\sigma \\
        0, & r\geq 2^{1/6}\sigma
    \end{cases} \label{eq:wca}
\end{equation}
where $r$ is the distance between pseudoatoms and $\sigma$ is the pseudoatom diameter, which we take as the unit of length for this model. Complementary attractor pseudoatoms interact via a Lennard-Jones potential:
\begin{equation}
    \uss(r) =  \begin{cases}  \ulj(r) - \ulj(\rcut) , & r < \rcut \\
    0, & r \geq \rcut
    \end{cases} 
    \label{eq:subsubbinding}
\end{equation}
where $\ulj$ is given by 
\begin{equation}
    \ulj(r) = 4\ebind \left[\left(\frac{\sigma}{r}\right)^{12}-\left(\frac{\sigma}{r}\right)^6\right]. 
    \label{eq:LennardJones}
\end{equation}
and $\rcut=3\sigma$. 
Two interacting subunits are depicted in Fig. 2E.

\subsection{Dynamics}

 Subunits (labeled $i=1,...,N$) evolve in time via Langevin dynamics:
\begin{subequations}
    \begin{align}
    m\frac{\dee\bv_i}{\dee t} &= \fcons[,i] -\gamma \bv_i + \frand[,i], \label{eq:langevin1}\\
    \bv_i &= \frac{\dee \br_i}{\dee t}, \label{eq:langevin2}\\
    \mathbf{I}\frac{\dee \bw_i}{\dee t} &= \tcons[,i] - \gamma_{\text{r}}\bw_i + \trand[,i] \label{eq:langevin3}\\
    \bw_i &= \frac{\dee \mathbf{q}_i}{\dee t}. \label{eq:langevin4}
    \end{align}
\end{subequations}
Here $m$ is the mass and $\mathbf{I}$ is the inertia tensor of a subunit; $\gamma$ and $\gamma_{\text{r}}$ are translational and rotational friction constants;  $\br_i$ and $\bv_i$ are the center of mass position and velocity; $\mathbf{q}_i$ and $\bw_i$ are the orientation and angular velocity; $\fcons[,i]$ and $\tcons[,i]$ are the conservative force and torque that come from the potential energy $U$, which includes a sum of pairwise interactions between pseudoatoms (Eqs. \ref{eq:uaa}, \ref{eq:utt}, and \ref{eq:utb} for the dodecahedron model; Eqs. \ref{eq:wca} and \ref{eq:subsubbinding} for the icosahedron model),  a sum over single-pseudoatom energies due to the condensate potential (Eq. \ref{eq:ucond})), and  constraint forces that ensure rigid body motion. The quantities $\frand[,i]$ and $\trand[,i]$ are the random (thermal) force and torque, which have zero mean and variances given by:
\begin{subequations}
    \begin{align}
        \langle \frand[,i](t) \cdot \frand[,j](t') \rangle &= 6\kt\gamma \delta(t-t')\delta_{ij} \\
        \langle \trand[,i](t) \cdot \trand[,j](t') \rangle &= 6 \kt \gamma_{\text{r}} \delta(t-t')\delta_{ij}. \label{eq:langevin}
    \end{align} 
\end{subequations} 
We set $\gamma=10$ and $\gammar=4\gamma/3$, which makes the dynamics effectively overdamped (consistent with the viscous interior of a cell). Within HOOMD-blue \cite{Anderson2020}, we use the velocity Verlet algorithm to solve Eqs. \ref{eq:langevin1}, \ref{eq:langevin2}, and the similarly symplectic integrator of Kamberaj et al. \cite{Kamberaj2005} to solve Eqs. \ref{eq:langevin3}, \ref{eq:langevin4}. We employ a timestep $\Delta t=5\times 10^{-3}$.

\section{Partition coefficient calculation}\label{supp_sec:partition}
\subsection{Ideal solution theory}
For sufficiently low concentrations of subunits within the condensate, ideal solution theory gives $\KcIS = \frac{\rho_1^{\text{c}}}{\rho_1^{\text{bg}}}=e^{\beta \Econd}$. For subunit densities and partition coefficients that lead to low packing fractions $\eta$ within the condensate $\eta \lesssim 10\%$ this theory predicts the measured partition coefficient with reasonable accuracy (Fig. 10 in main text). For larger values, we observe significant deviations between ideal solution theory and simulation results without assembly, in which we set $\ebind=0$ for attractor pseudoatoms (but keep $\ebind=1$ for icosahedron model excluder pseudoatoms). 

\subsection{Hard sphere theory}
To account for subunit excluded volume while computing $\Kc$, we model subunits as effective hard spheres with diameter $\sigone$. At equilibrium, subunits in the condensate and background must have equal chemical potentials:
\begin{equation}
    \muc = \mubg. \label{eq:equal_mu_noassem}
\end{equation}
The chemical potentials can be written in terms of ideal and excess (hard sphere) components:
\begin{subequations}
    \begin{align}
        \mubg &= \muex(\rhoonebg\sigone^3) + \kt \log{(\rhoonebg\sigone^3)} \label{eq:mubg} \\
        \muc &= \muex(\rhoonec\sigone^3) + \kt \log{(\rhoonec\sigone^3)} - \Econd. \label{eq:muc}
    \end{align}
\end{subequations}
Within the Carnahan-Starling approximation, $\muex$ is given by \cite{Attard1993}:
\begin{equation}
    \muex(\rho\sigone^3) = \frac{\eta(3\eta^2-9\eta+8)}{(1-\eta)^3}, \;\;\; \eta = \frac{\pi}{6}\rho\sigone^3,\label{eq:carnahan-starling}
\end{equation}
where $\eta$ is the hard sphere packing fraction.  Plugging Eqs. \ref{eq:mubg}, \ref{eq:muc} in Eq. \ref{eq:equal_mu_noassem} and rearranging, we obtain:
\begin{equation}
    \frac{\rhoonec}{\rhoonebg} = e^{\beta\left[\Econd + \muex(\rhoonebg\sigone^3) - \muex(\rhoonec\sigone^3)\right]}. \label{eq:rho_self_consistent}
\end{equation}
Combined with the mass conservation condition
\begin{equation}
    \rhoonebg = (1+\Vr)\rhoT - \Vr \rhoonec ,
\end{equation}
Eq. \ref{eq:rho_self_consistent} can be solved self-consistently to obtain $\rhoonec$ and, hence, $\Kc$. We set $\sigone=2.1$ (the `T' pseudoatom diameter) for the dodecahedron subunits and $\sigone=3$ (the side length) for the icosahedron subunits. 
As shown in Fig. 10, we obtain excellent agreement between this hard sphere approximation and results from simulations with $\ebind=0$ over a wide range of $\Econd$. In contrast, ideal solution theory breaks down for $\Econd\gtrsim4$ (with $\rhoT=4\times 10^{-4}$), corresponding to a packing fraction of $\approx 10\%$.

\section{Equilibrium assembly theory}\label{supp_sec:hs_assembly_theory}

\subsection{Ideal solution theory}

Here, we summarize an equilibrium theory for assembly coupled to LLPS in the ideal solution limit (in which excluded volume is neglected). This theory was previously developed in Ref.~\cite{Hagan2023}, and a similar theory was presented in Ref.~\cite{Bartolucci2024}. At equilibrium, intermediates are vanishingly rare \cite{Hagan2021} and thus we assume that subunits are  either monomers or part of complete capsids containing $\ncap$ subunits ($\ncap=12$ for dodecahedral capsids and $\ncap=20$ for icosahedral capsids). Both monomers and capsids can be located  in the condensate or in the background. We denote the concentrations of capsids in the condensate and background as $\rhonc$, $\rhonbg$ respectively. Equilibrium implies that, within each phase (condensate and background), the monomer  and (per subunit) capsid chemical potentials are equal, and  that monomer and capsid chemical potentials are equal between phases:

\begin{subequations}\label{eq:equal_mu}
    \begin{align}
    \muonebg &= \munbg \\
    \muonec &= \munc \\
    \muonebg &= \muonec \\
    \munbg &= \munc, 
\end{align}
\end{subequations}
with $\muonebg$, $\muonec$, $\munbg$, and $\munc$ as the monomer and (per subunit) capsid chemical potentials in the background and condensate, respectively. Within the ideal solution approximation, the chemical potentials are given by \cite{N.Israelachvili1976}:
\begin{subequations}\label{eq:mu_def}
\begin{align}
    \muonebg &= \kt\log{(\rhoonebg\sigone^3)} \\
    \muonec &= \kt\log{(\rhoonec\sigone^3)} - \Econd \\
    \munbg &= \kt\log{(\rhonbg\sigone^3)}/\ncap + \fcapsid \\
    \munc &= \kt\log{(\rhonc\sigone^3)}/\ncap + \fcapsid - \Econd. 
\end{align}
\end{subequations}
with $\fcapsid$ as the per-subunit Helmholtz free energy of capsid formation. Additionally, mass conservation yields the constraint:
\begin{equation}
    (1+\Vr)\rhoT = \Vr \rhoonec + \rhoonebg + \ncap\Vr \rhonc + \ncap\rhonbg. \label{eq:mass_cons}
\end{equation}
We solve Eqs.~\ref{eq:equal_mu} and \ref{eq:mass_cons} self-consistently (see Mathematica Notebook 1 available on the Open Science Framework
OSFHome: https://osf.io/hq2y8/) to yield the concentrations of monomers and capsids in the background and condensate, and hence the total equilibrium yield:
\begin{equation}
    \fcequil = \frac{\ncap}{\rhoT} \left(\frac{\Vr}{1+\Vr}\rhonc + \frac{1}{1+\Vr}\rhonbg\right). 
\end{equation}
Following Ref.~\cite{Hagan2023}, we note that in the limit of large capsid size ($N\gg1$), the equilibrium yield can be written as:
\begin{equation}
    \fcequil = \begin{cases}
        1-\frac{\rhocac}{\rhoT}, &\rhoT\gg \rhocac \\
        \left(\frac{\rhoT}{\rhocac}\right)^N, & \rhoT \ll \rhocac,
    \end{cases}
\end{equation}
with  the critical assembly concentration:
\begin{equation}
    \rhocac = \frac{1+\Kc\Vr}{\Kc(1+\Vr)}\left(\frac{1+\Vr}{\Vr}\right)^{1/N}\rhocac^0,
\end{equation}
and $\rhocac^0$ as the critical assembly concentration without LLPS:
\begin{equation}
    \rhocac^0 = N^{-1/N}e^{\fcapsid/\kt}.
\end{equation}
Ideal solution theory thus predicts that increasing $\Kc$ and decreasing $\Vr$ reduce $\rhocac$. For the dodecahedron model, we set $\fcapsid=-5\kt$ for $\ebind=6$ (see SI Section~\ref{supp_sec:param_estimate}B for details on this estimate).

\subsection{Hard sphere theory}

We now extend ideal solution theory to account for excluded volume. As in the ideal solution theory, at equilibrium we will have equality of chemical potentials between phases and between monomers and capsids within each phase (Eq.~\ref{eq:equal_mu_noassem}). To account for excluded volume, we model monomers and capsids as hard spheres with diameters $\sigone$ and $\sigcap$, respectively. The chemical potentials are now given by:
\begin{subequations}\label{eq:mu_def_hard}
\begin{align}
    \muonebg &= \kt\log{(\rhoonebg\sigone^3)} + \muexone(\rhoonebg, \rhonbg) \\
    \muonec &= \kt\log{(\rhoonec\sigone^3)} + \muexone(\rhoonec, \rhonc) - \Econd \\
    \munbg &= \kt\log{(\rhonbg\sigone^3)}/\ncap + \muexcap(\rhoonebg, \rhonbg)/\ncap + \fcapsid \\
    \munc &= \kt\log{(\rhonc\sigone^3)}/\ncap + \muexcap(\rhoonec, \rhonc)/\ncap + \fcapsid - \Econd. 
\end{align}
\end{subequations}
Here, $\muexone$ and $\muexcap$ are excess chemical potentials for a binary mixture of hard spheres~\cite{Mansoori1971,Heyes2016}:
\begin{subequations}
\begin{align}
    \muexnu(\rhoone,\rhon) &= c_0(\eta) + c_1(\eta)\frac{M_1M_2}{M_3}\frac{\sigone}{M_1} + \left(c_1(\eta)\frac{M_1M_2}{M_3}+3a_2(\eta)\frac{M_2^3}{M_3^2}\right)\frac{\sigone^2}{M_2} \nonumber \\
    &+ \left[\eta c_0'(\eta) + \left(\eta c_1'(\eta)-c_1(\eta)\right)\frac{M_1M_2}{M_3} + \left(\eta a_2'(\eta)-2a_2(\eta)\right)\frac{M_2^3}{M_3^2}\right]\frac{\sigone^3}{M_3} \\
    \eta &= \frac{\pi}{6}(\rhoone+\rhon)M_3 \\
    c_0(\eta) &= -\log{(1-\eta)} \\
    c_1(\eta) &= \frac{3\eta }{1-\eta} \\
    a_2(\eta) &= \log{(1-\eta)} + \frac{\eta}{(1-\eta)^2} \\
    M_i &= \frac{\sum_{\nu}\rho_{\nu}\sigma_{\nu}^i}{\sum_{\nu}\rho_{\nu}},
\end{align}
\end{subequations}
where $\nu=1,\text{cap}$ and the prime symbol denotes a derivative with respect to $\eta$.

\begin{figure}[ht]
    \centering
    \includegraphics[width=0.5\linewidth]{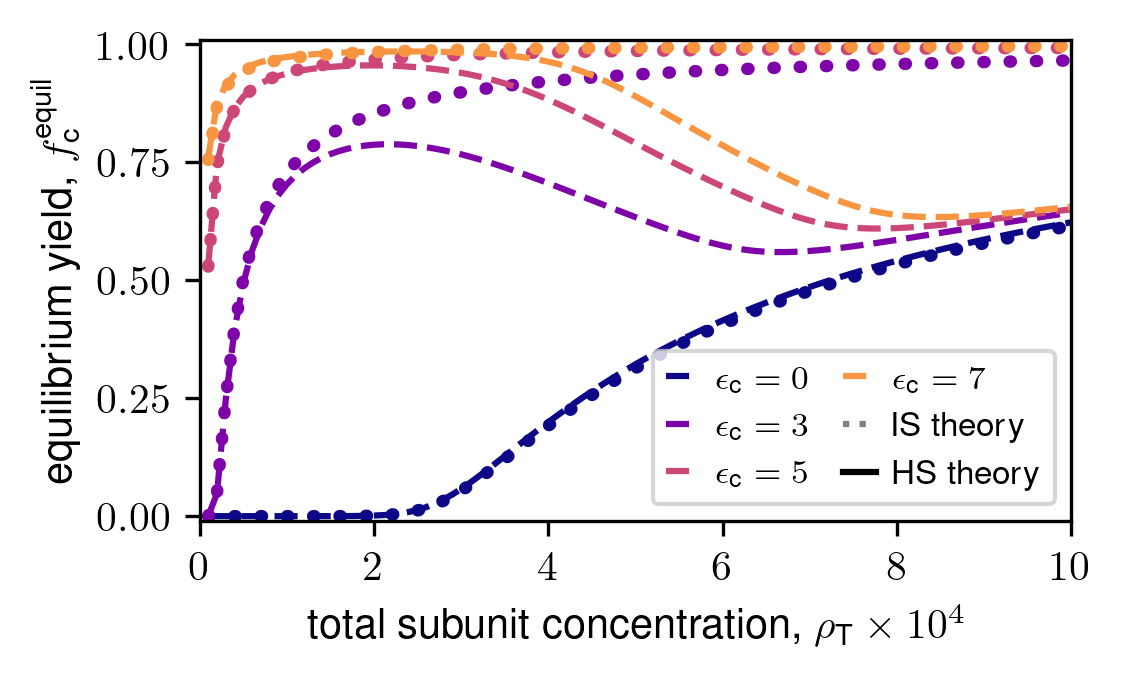}
    \caption{Equilibrium yield versus total subunit concentration, predicted by the HS (solid lines) and IS (dotted lines) theories for indicated values of $\Econd$. Parameters are: the subunit binding well depth $\ebind =6$, and condensate volume fraction $\Vr=5.0\times 10^{-3}$, and final simulation time $\tF=6\times 10^5$.}
    \label{supp_fig:equil_yield_vs_conc}
\end{figure}

We solve Eqs.~\ref{eq:equal_mu} and \ref{eq:mass_cons} using the chemical potentials in Eq.~\ref{eq:mu_def_hard} (see Mathematica Notebook 2 available on the Open Science Framework
OSFHome: https://osf.io/hq2y8/) to obtain the equilibrium concentrations of monomers and capsids in each phase, and hence the equilibrium yield. For the dodecahedron model, we set $\sigone=2.1$ (the `T' pseudoatom diameter), $\sigcap=5.1$, and $\fcapsid=-5\kt$ (see SI Section~\ref{supp_sec:param_estimate}A for details on estimating $\sigcap$). We plot both the ideal solution and hard sphere predictions for $\fcequil$ versus $\rhoT$ for different values of $\Econd$ in Fig.~\ref{supp_fig:equil_yield_vs_conc}. For $\Econd=0$ the ideal solution and hard sphere results closely match, as expected for a dilute solution. On the other hand, when a condensate is present ($\Econd>0$), the hard sphere predictions deviate significantly from ideal solution theory beyond a threshold concentration, reflecting the high concentration of subunits within the condensate.

\section{Estimating parameters for the theoretical model}\label{supp_sec:param_estimate}

\subsection{Estimating $\sigcap$}

\begin{figure}
    \centering
    \includegraphics[width=0.5\linewidth]{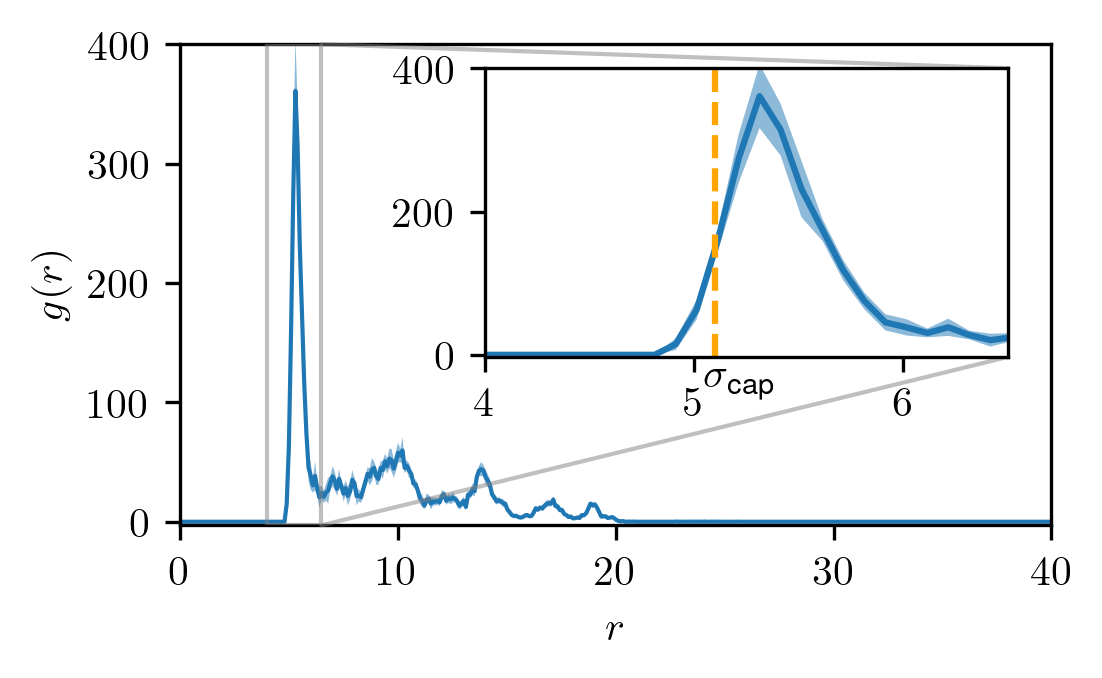}
    \caption{Capsid radial distribution function $g(r)$ versus center-of-mass distance $r$, computed from simulations with $\Vr=5.0\times10^{-3}$, $\rhoT=10^{-3}$, $\ebind=6$, and $\Econd=7$. Here $g(r)$ was computed by averaging over the final configurations of 10 independent assembly trajectories, with error bars representing the standard error. The inset shows a zoomed-in view of the first peak, and the orange dashed line shows the capsid effective hard sphere diameter $\sigcap$.}
    \label{supp_fig:gr}
\end{figure}

To estimate the capsid effective hard sphere diameter, we computed the capsid radial distribution function $g(r)$ (Fig.~\ref{supp_fig:gr}) at a parameter set where condensates are densely packed with capsids ($\rhoT=10^{-3}$, $\Vr=5.0\times 10^{-3}$, $\ebind=6$, $\tF=6\times10^5$). We took $\sigcap$ to be halfway between the location of the first peak of $g(r)$ ($\approx 5.3$) and the smallest value of $r$ for which $g(r)$ is nonzero ($\approx 4.9$), giving an estimate of $\sigcap\approx5.1$. We also estimated $\sigcap$ via an alternative route, in which we considered assembled capsids as rigid bodies and computed the minimum distance that two capsids could approach each other. In snapshots of densely packed capsids, we often observed capsids touching either face-to-face or vertex-to-face (with a `T' pseudoatom on one capsid nestled between a face consisting of three `T' pseudoatoms on another capsid). We therefore computed the distance of closest approach in either of these two orientations, obtaining values of $\approx4.8$ for face-to-face and $\approx5.3$ for vertex-to-face (see Mathematica Notebook 3 available on the Open Science Framework
OSFHome: https://osf.io/hq2y8/). The average of these two values is approximately equal to our previous estimate of $\sigcap\approx 5.1$. 

\subsection{Estimating $\fcapsid$}

\begin{figure}[ht]
    \centering
    \includegraphics[width=0.5\linewidth]{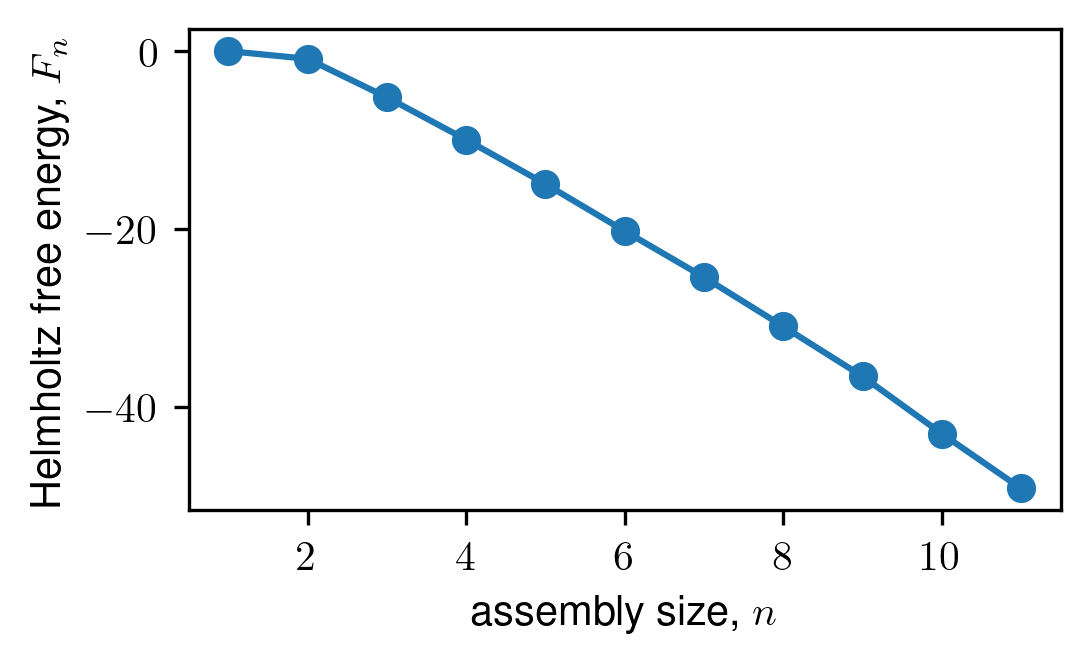}
    \caption{Helmholtz free energy $F_n$ as a function of assembly size $n$. Data taken from Ref.~\cite{Trubiano2024a}.}
    \label{supp_fig:capsid_free_energy}
\end{figure}

To estimate $\fcapsid$, we used previously reported approximate free energies for capsid intermediates up to, but not including, a full capsid in the dodecahedron model (see Ref.~\cite{Trubiano2024}). We plot these free energies for $\ebind=6$ in Fig.~\ref{supp_fig:capsid_free_energy}. 
For intermediate sizes $n\gtrsim 3$, the free energy to add an additional subunit is roughly independent of $n$, $\approx -5.5 \kt$, until the 11th subunit binds. The accuracy of these free energy estimates is limited by sampling for larger intermediates and we would expect a more nonlinear dependence on $n$ (since more bonds are formed on average as intermediates grow and the binding entropy is expected to be sub-linear in number of bonds \cite{Hagan2006, Hagan2011}). Furthermore, because the insertion of the 12th subunit (making a complete capsid) makes more bonds (5) compared to previous subunits than any other, this reaction is irreversible on simulation timescales. That is, we never see a complete capsid disassemble at binding energy values for which assembly occurs. Given the irreversibility and larger number of bonds formed, we anticipate that the free energy of adding the final subunit is significantly higher than that of adding previous subunits. We thus roughly approximate the free energy of adding the final subunit as $-11 \kt$, roughly twice that of adding previous subunits, such that the total capsid free energy is $\approx -60\kt$. Thus, the free energy per subunit is $\fcapsid\approx -5\kt$. However, the capsid free energy will depend on $\ebind$. To estimate this dependence, we use the previously-reported~\cite{Trubiano2024a} dependence of dimerization free energy on $\ebind$, $\sim -1.56\ebind + Ts$, where $s$ is the per-subunit entropy. Assuming that the capsid free energy depends on $\ebind$ in the same way as the dimerization free energy, we thus write:
\begin{equation}
    \fcapsid(\ebind) = -1.56\ebind + Ts.
\end{equation}
We then use the fact that $\fcapsid(6\kt)=-5\kt$ to obtain $Ts=4.36\kt$. 

\section{Yield vs $\ebind$ at different times}\label{supp_sec:yield_time_series}

\begin{figure}[ht]
    \centering
    \includegraphics[width=1.0\linewidth]{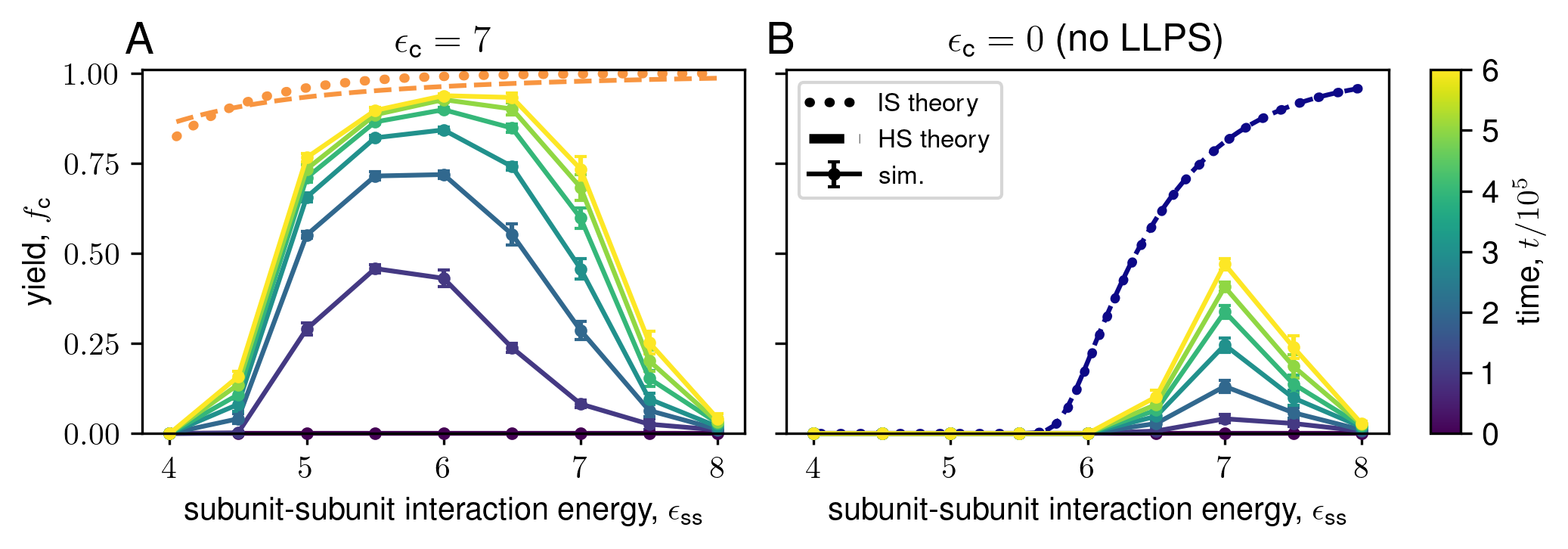}
    \caption{ Yield as a function of $\ebind$ at indicated times for  (A) $\Econd=7$ and (B) $\Econd=0$ (i.e. without LLPS), for the same parameters as Fig. 6A.
 The equilibrium hard sphere and ideal solution theories are shown as dashed and dotted lines respectively. }
    \label{supp_fig:yield_vs_ess_time_series}
\end{figure}

Fig.~\ref{supp_fig:yield_vs_ess_time_series}a,b plots the yields from Fig. 6 at different times for $\Econd=7$ and $\Econd=0$ respectively. With LLPS, capsid production is orders of magnitude faster and yields approach much closer to the estimated equilibrium value within the finite simulation timescale. Moreover, assembly rates are increased even in conditions leading to malformed assemblies ($\ebind \ge 7.5$). However, consistent with the results described in the main text, assembly growth slows at later times. While the slowdown with LLPS primarily occurs due to the excluded volume effects described above, assembly rates in bulk solution decrease as monomers are depleted, leading to a growth in the size of the critical nucleus size and the associated free energy barrier \cite{Hagan2010}.

\section{Scaling theory for assembly kinetics}\label{supp_sec:scaling_theory}

The kinetics of LLPS-mediated assembly involves diffusion of subunits into the condensate, formation of critical nuclei, and elongation (growth) of nuclei into complete capsids. Previous work~\cite{Hagan2023} predicted how LLPS affects the maximum assembly rate and median assembly timescale by noting that nucleation is typically the rate-limiting step in self-assembly, and by assuming that the primary effect of LLPS is to increase the nucleation rate by increasing the local subunit concentration. Also assuming that the critical nucleus size $\nNuc$ is independent of conditions, 
 the maximum assembly rate $\rMax$ and median assembly time $\tHalf$ were predicted to depend on the partition coefficient $\Kc$ and condensate volume fraction $\Vr$ as:
\begin{align}
\rMax(\Vr,\Kc) \approx & \rNuc(\Vr,\Kc) =  \sNuc \rNucND \nonumber \\
\tHalf(\Vr,\Kc) = & \tHalf^0/\sNuc  \nonumber \\
\sNuc = & \left(\frac{1+\Vr}{1+\Kc\Vr}\right)^{\nNuc}\frac{1+\Vr\Kc^{\nNuc}}{1+\Vr} \nonumber\\
\approx & \Vr / \left(\Vr + 1/\Kd\right)^{\nNuc}.
\label{eq:scaling}
\end{align}
Here, $\rNucND$ and $\tHalfND$ are the initial nucleation rate (before significant subunit depletion has occurred) and median assembly time in the absence of LLPS ($\Econd=0$). The approximate expression for the nucleation speedup $\sNuc$ is valid for $\Kc^{\nNuc}\gg1/\Vr$, such that nucleation occurs only in the compartment.

This previously-developed scaling theory assumes diffusion is very fast compared to nucleation. Yet this assumption is violated in our simulations, particularly for high $\Econd$, as shown in Fig. 7C,F in the main text. We therefore extend the scaling theory to account for diffusive timescales. First, we note that the maximum assembly rate will be limited by the diffusive flux of subunits into the condensate, (assuming that assembly occurs exclusively within the condensate) with a ``forward'' rate (i.e. diffusion \textit{into} the condensate) given by: $4\pi \Rc D \rhoonebg$ ($\approx 4\pi \Rc D \rhoT$, assuming minimal subunit depletion). Second, we note that the median assembly time must account for the characteristic timescale for subunit diffusion between condensate and background (not just the forward flux of subunits into the compartment). To obtain this timescale, we follow the Supporting Material of Ref.~\cite{Hagan2023} and write the dynamics of subunit partitioning between the background and compartment as:
\begin{subequations}
\begin{align}
        \frac{\dee \rhoonebg}{\dee t} &= -\frac{4\pi \Rc D}{\Vc}\left(\rhoonebg - \rhoonec/\Kc\right) \\
        \frac{\dee \rhoonebg}{\dee t} &= -\frac{\Vc}{\Vbg}\frac{\dee \rhoonec}{\dee t}.
\end{align}
\end{subequations}
Assuming initial concentrations $\rhoT$ in both the background and condensate results in:
\begin{equation}
    \rhoonec(t) = \left(\frac{1+\Vr}{\Vr+1/\Kc}\right)\rhoT + \left(1-\frac{1+\Vr}{\Vr+1/\Kc}\right)\rhoT e^{-(4\pi \Rc D (\Vr+\Kc^{-1})/\Vc)t}
\end{equation}
We identify $\Vc/\left[(\Vr+\Kc^{-1})4\pi \Rc D\right]$ as a diffusive timescale. However, this expression fails to account for the subunits that will form capsids. To roughly capture the diffusion timescale associated with those subunits, we add a term $\fc V/(4\pi \Rc D)$, resulting in a total diffusive timescale:
\begin{equation}
    \tDiff \approx \frac{\Vc/(\Vr+1/\Kc) + \fc V}{4\pi \Rc D},
\end{equation}
which is Eq. 5 in the main text. Thus, our new scaling estimates for $\rMax$ and $\tHalf$ are:
\begin{subequations}
    \label{eq:scaling_new}
    \begin{align}
        \rMax(\Vr,\Kc) = & \frac{\sNuc\rNucND 4\pi \Rc D \rhoT}{\sNuc\rNucND + 4\pi \Rc D \rhoT} \frac{\rhoT}{\ncap}\\
        \tHalf(\Vr,\Kc) = & \tHalfND/\sNuc + \tDiff/2.
    \end{align}
\end{subequations}
(We have included the factor of $\rhoT/\ncap$ to give $\rMax$ units of capsid concentration per unit time.)

\begin{figure}
    \centering
    \includegraphics[width=1.0\linewidth]{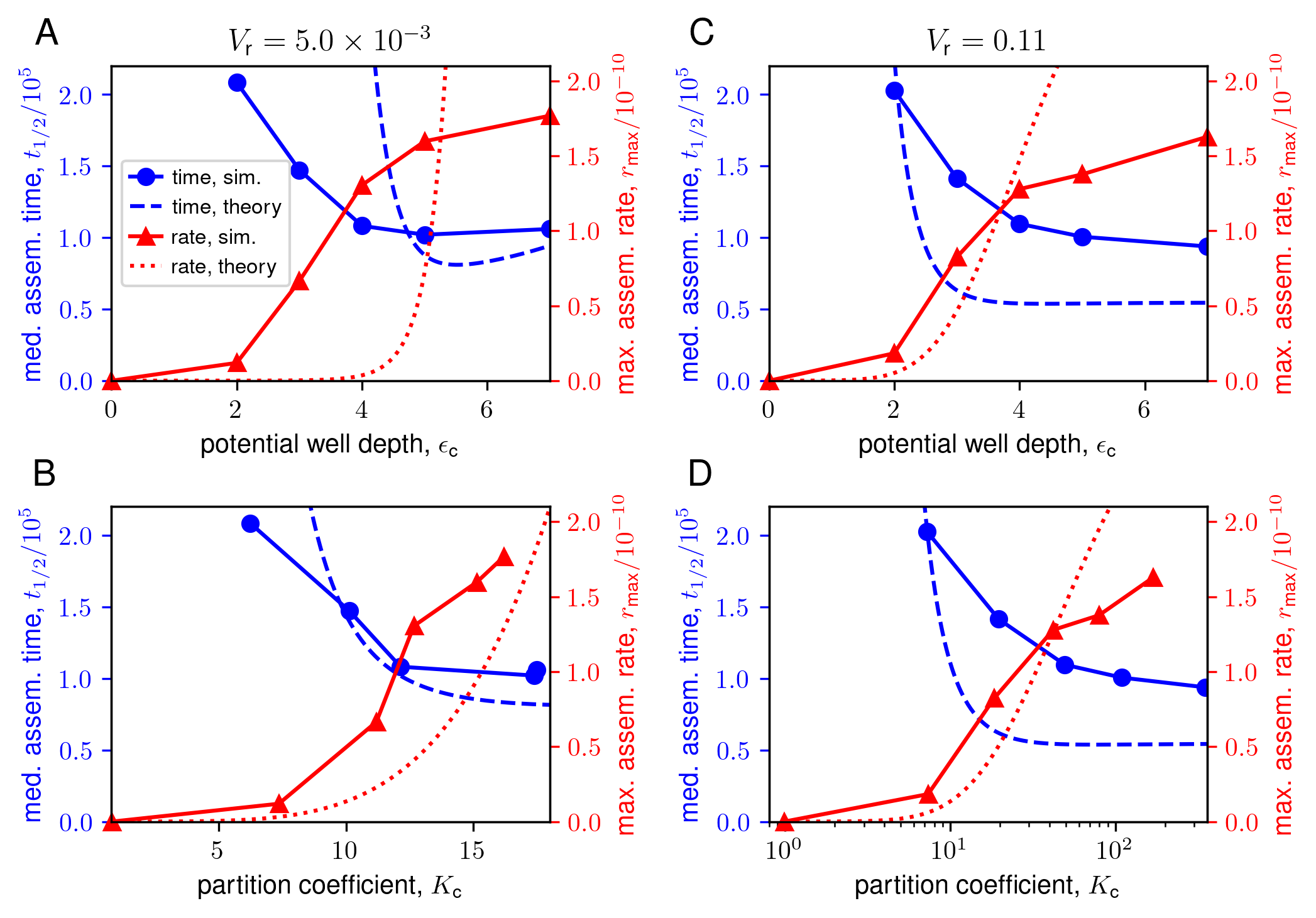}
    \caption{Comparison of simulations and scaling theory for median assembly time $\tHalf$ and maximum assembly rate $\rMax$. Panels A,B show results for $\Vr=5.0\times10^{-3}$, and panels C,D show results for $\Vr=0.11$. Panels A,C show results as a function of potential well depth $\Econd$, while panels B,D show the same results as a function of partition coefficient $\Kc$. The scaling theory curves (Eq.~\ref{eq:scaling}) were fit by eye to the simulation data for $\Vr=0.11$ as a function of $\Econd$, resulting in fitting parameters $\rNucND=10^{-9}$, $\tHalf^0=3\times 10^7$, and $\nNuc=5$.}
    \label{supp_fig:rate_time_theory_vs_sim}
\end{figure}

We compare the theoretical predictions of Eq.~\ref{eq:scaling_new} to simulation results in Fig.~\ref{supp_fig:rate_time_theory_vs_sim}. Because we observe no assembly without LLPS for the parameters used in Fig. 7, we treat $\rNucND$ and $\tHalfND$, as well as $\nNuc$, as fitting parameters. We choose (by eye) $\rNucND=10^{-9}$, $\tHalf^0=3\times 10^7$, and $\nNuc=5$ to yield rough agreement with the simulation results for $\Vr=0.11$, and then use the same parameter values for $\Vr=5\times10^{-3}$. We find that the diffusive flux term $4\pi \Rc D\rhoT$ has little effect on $\rMax$ because here  $4\pi \Rc D\rhoT\gg \sNuc\rNucND$; however, we note that the true diffusive flux could be significantly smaller due to the excluded volume of subunits and capsids in the condensate ``blocking'' entry of additional subunits. However, including $\tDiff/2$ in $\tHalf$ correctly captures the saturation of $\tHalf$ with $\Econd$ (without $\tDiff$, the theory severely underestimates the median assembly time at high $\Econd$). 

When we assume the ideal solution partition coefficient $(\KcIS=e^{\beta \Econd})$ and plot the scaling prediction as a function of $\Econd$ (Fig.~\ref{supp_fig:rate_time_theory_vs_sim}A,C), we observe poor agreement, which worsens with increasing $\Econd$, particularly for $\rMax$. However, we observe much better agreement when we instead use the values of $\KcActual$ measured at the time of maximum assembly rate for $\rMax$ and at the median assembly time for $\tHalf$ (Fig.~\ref{supp_fig:rate_time_theory_vs_sim}B,D). The agreement particularly improves for $\Vr=5\times10^{-3}$,  where the $\KcActual$ values deviate significantly from $\KcIS$.

\section{Additional supplementary figures}

\begin{figure}
    \centering
    \includegraphics[width=0.5\linewidth]{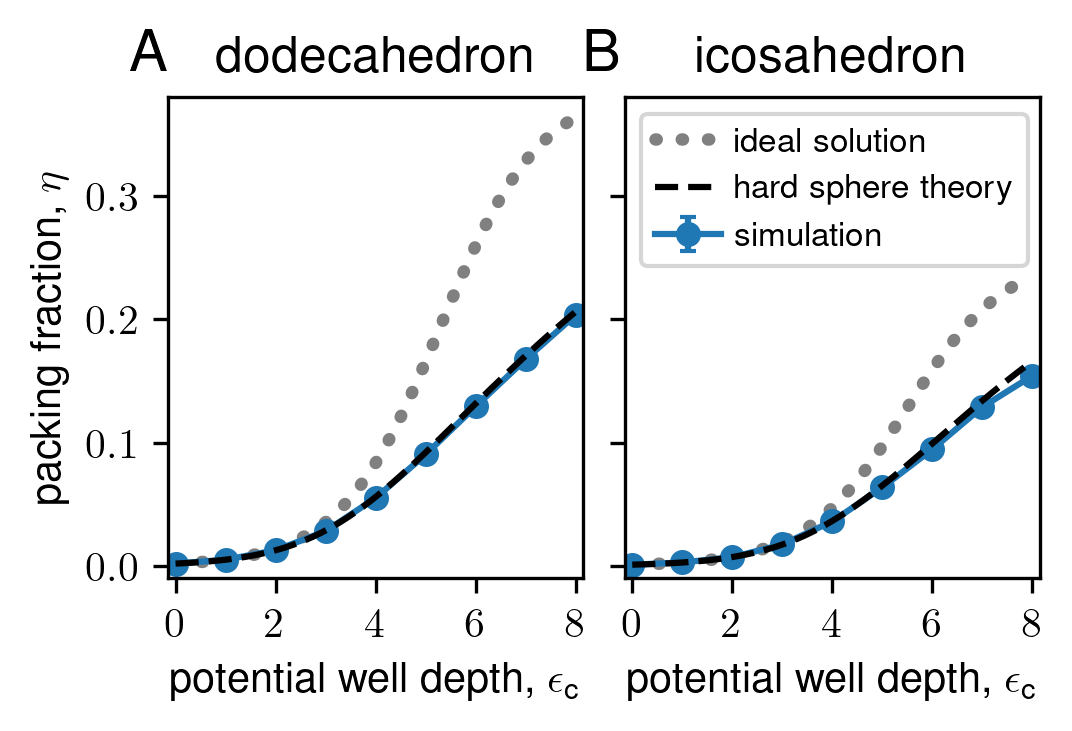}
    \caption{Packing fraction in the condensate $\eta=\frac{4}{3}\pi(\sigone/2)^3\rhoonec$ 
    as a function of potential well-depth $\Econd$ for the (A) dodecahedron model and (B)  icosahedron model. Error bars are smaller than the size of the plotted points.}
    \label{supp_fig:eta_vs_ec}
\end{figure}

\begin{figure}
    \centering
    \includegraphics[width=0.5\linewidth]{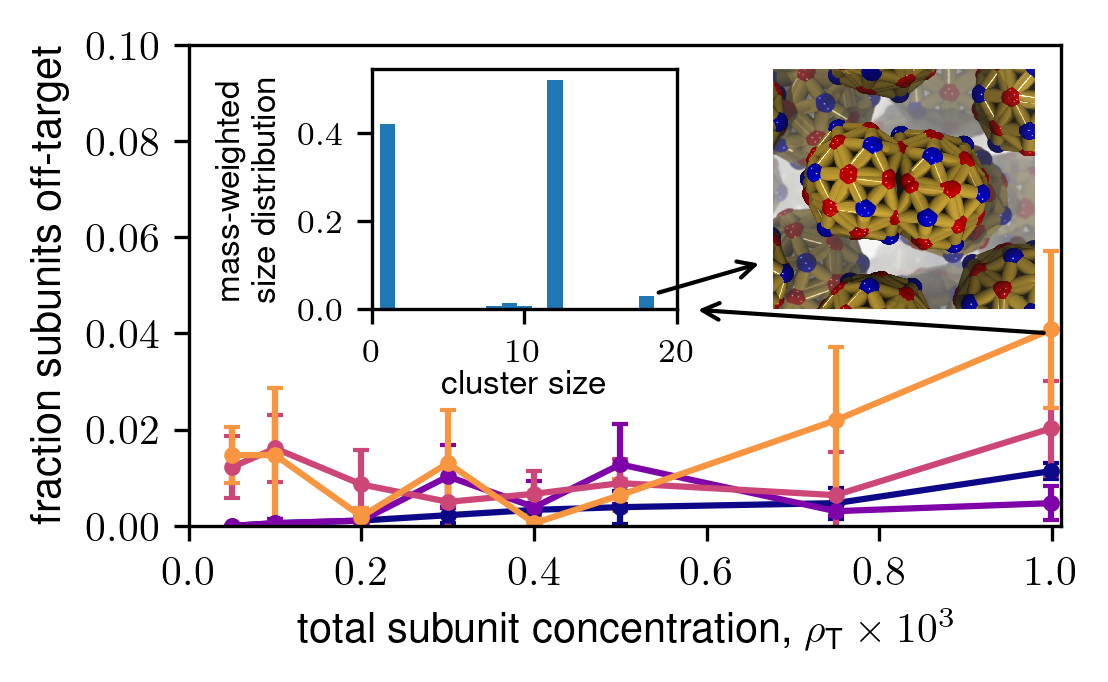}
    \caption{Fraction of subunits in off-target structures (neither monomers nor capsids) versus total subunit concentration for different partition coefficients. The inset plot shows the mass-weighted cluster size distribution for $\rhoT =10^{-3}$, $\Econd=-7\kt$. The two largest peaks at cluster sizes of 1 and 12 correspond to monomers and capsids, respectively. The next largest peak is at 18, corresponding to partial capsids bound together as shown in the snapshot.}
    \label{supp_fig:frac_off_target_vs_conc}
\end{figure}

\begin{figure}[ht]
    \centering
    \includegraphics[width=0.5\linewidth]{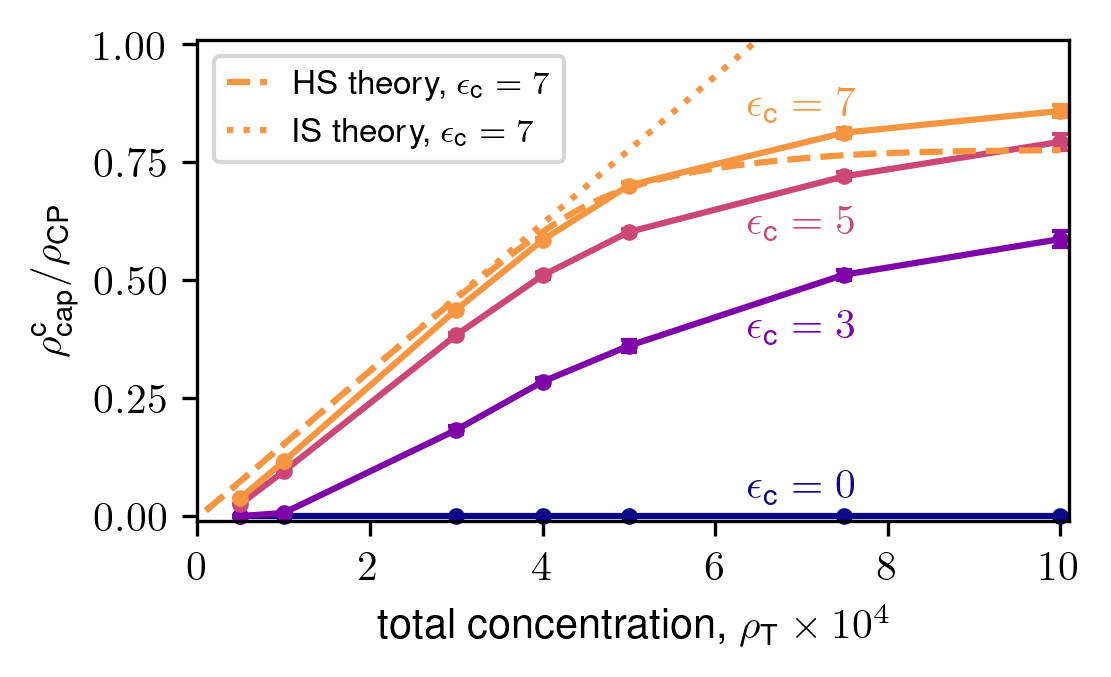}
    \caption{Concentration of capsids in the condensate normalized by the estimated close-packing density, $\rhonD/\rhocp$, versus total subunit concentration for the same parameters as in  Fig. 3.  }
    \label{supp_fig:capsid_conc_vs_rhoT}
\end{figure}

\begin{figure}[ht]
    \centering
    \includegraphics[width=0.5\linewidth]{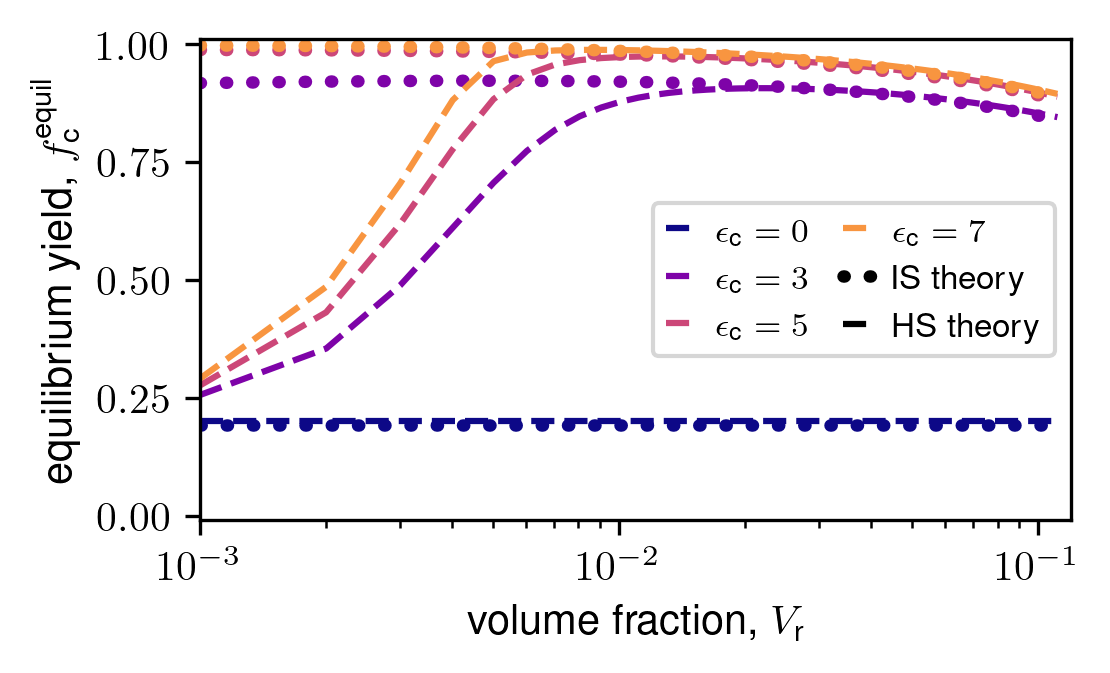}
    \caption{Equilibrium yield versus volume fraction for indicated partition coefficient parameters, shown for the  IS (dots) and HS (solid) theories. Parameters are $\ebind =6$, $\rhoT=4.00\times 10^{-4}$, and $\tF=6\times 10^5$.}
    \label{supp_fig:equil_yield_vs_vr}
\end{figure}

\begin{figure}
    \centering
    \includegraphics[width=0.5\linewidth]{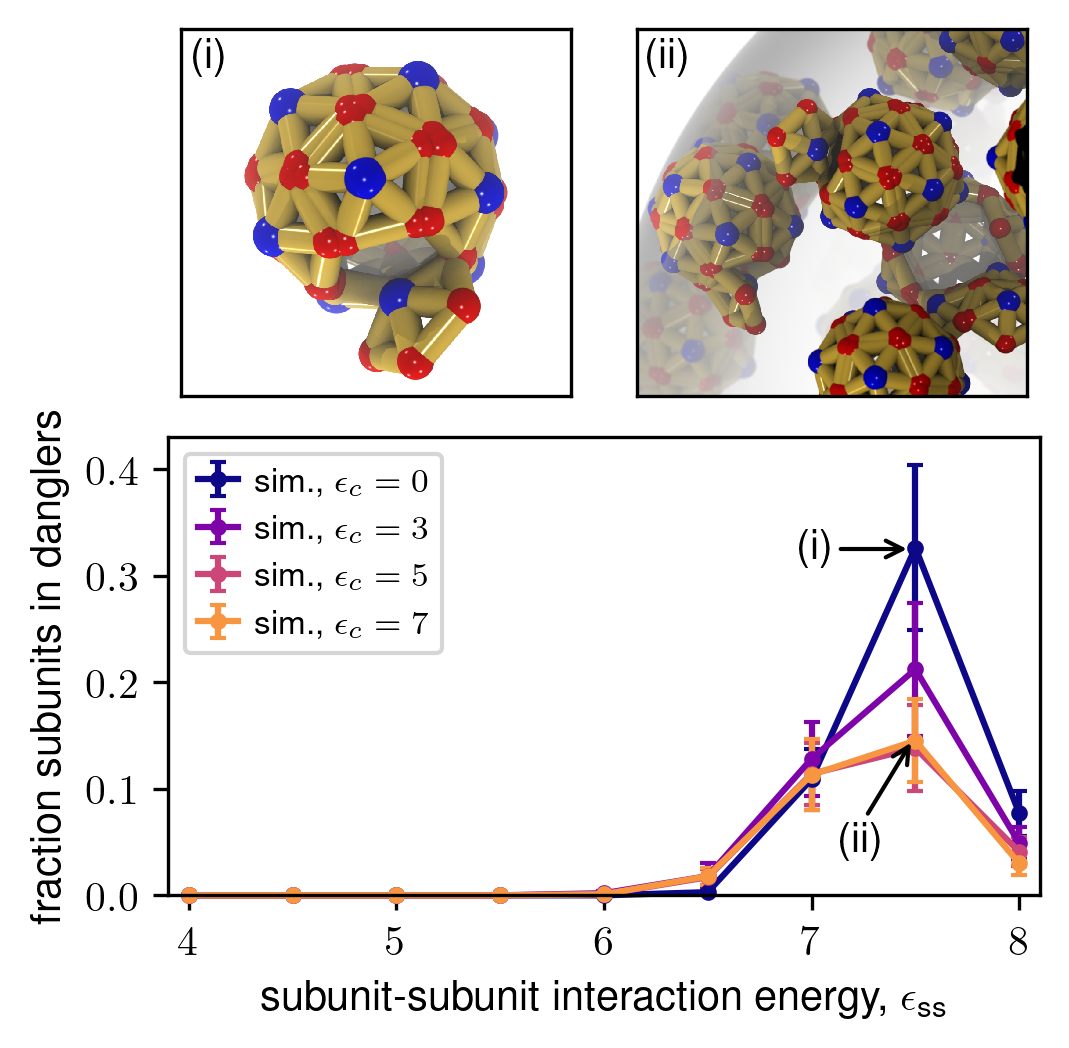}
    \caption{Fraction of subunits in danglers versus subunit-subunit interaction energy. Snapshots above plot correspond to labeled points: (i) dangler in bulk solution with $\Econd=0$ (ii) danglers in a condensate with $\Econd=7$. Parameters are: $\rhoT=4\times10^{-4}$, $\Vr = 5.0 \times 10^{-3}$, and $\tF=10^6$}
    \label{supp_fig:danglers}
\end{figure}

\begin{figure}
    \centering
    \includegraphics[width=0.5\linewidth]{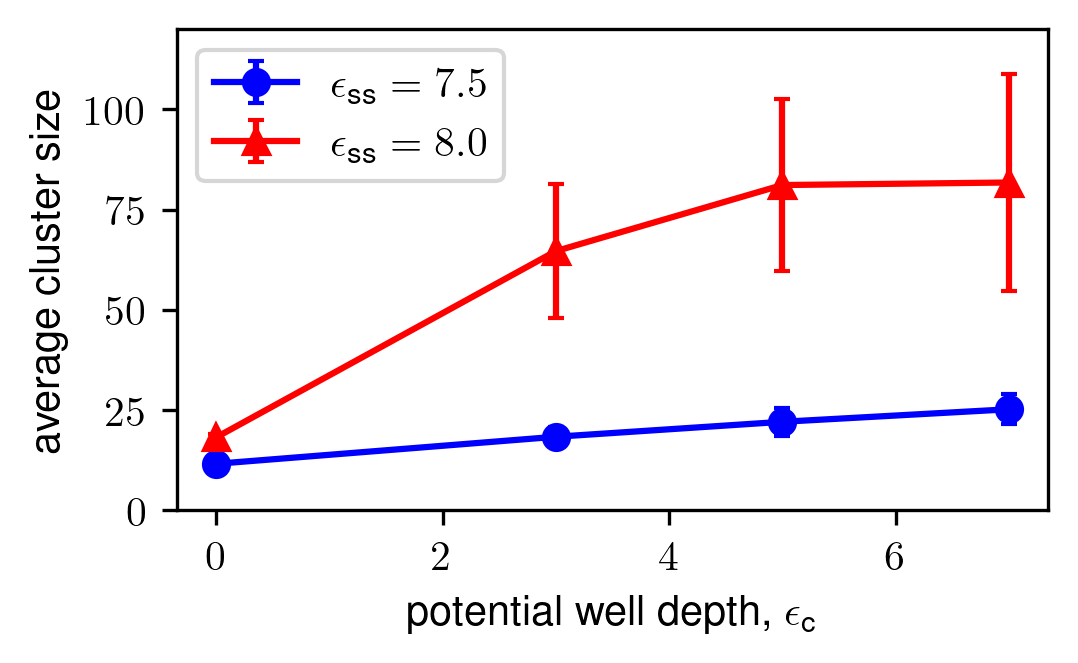}
    \caption{The average cluster size at long times is shown as a function of $\Econd$ for $\ebind=7.5,8$. Parameters are: $\rhoT=4\times 10^{-4}$, $\Vr=5.0\times 10^{-3}$, $\tF=6\times 10^5$.}
    \label{supp_fig:cluster_size}
\end{figure}

\begin{figure}
    \centering
    \includegraphics[width=0.5\linewidth]{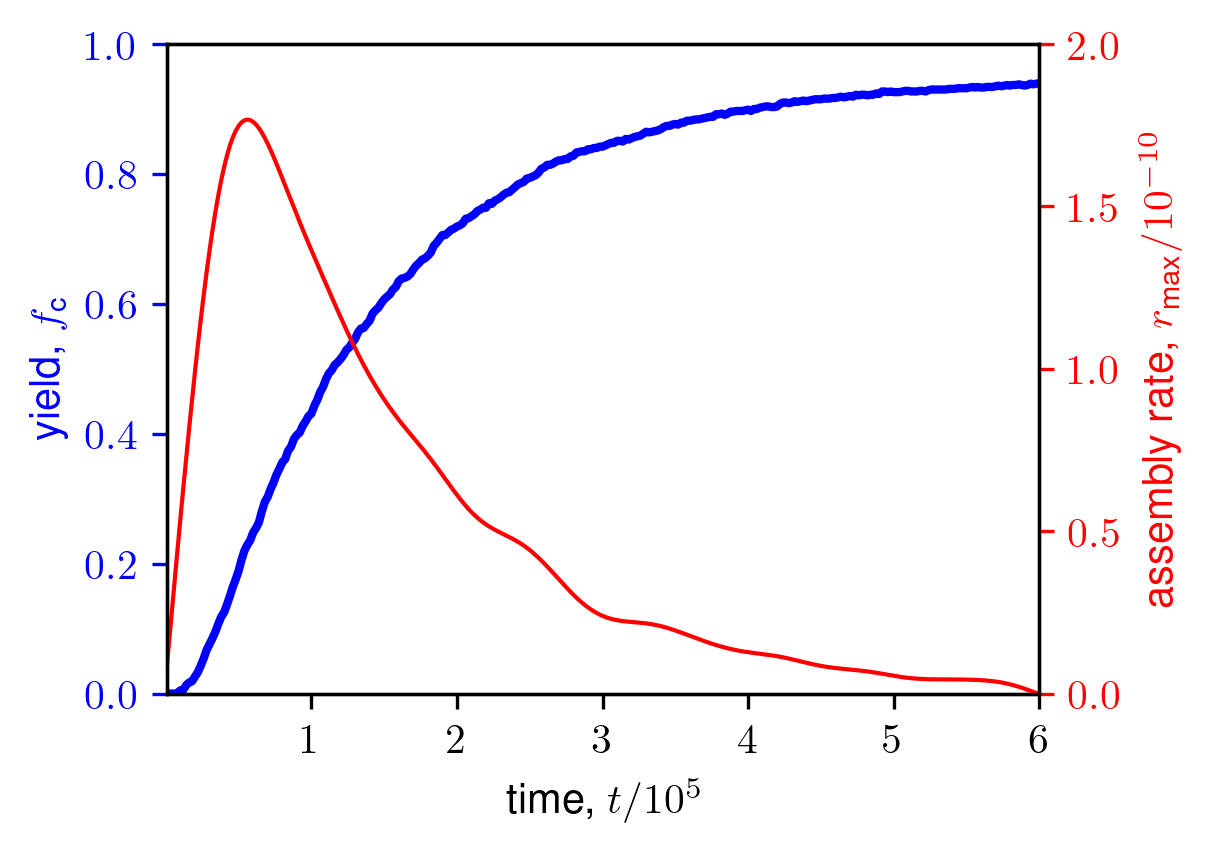}
    \caption{Example of assembly rate calculation from yield vs. time data. The yield is the thick blue line; the assembly rate is the thin red line. The assembly rate is computed by convolving the yield curve with the derivative of a Gaussian; we set the smoothing parameter (i.e. the standard deviation of the Gaussian) to $\sigma=10$. Simulation parameters are: $\Econd=7$, $\ebind=6$, $\Vr=5.0\times10^{-3}$, $\rhoT=4\times 10^{-4}$.}
    \label{supp_fig:assem_rate_fit}
\end{figure}

\clearpage

\section{Movie descriptions}
\begin{itemize}

    \item \textbf{Movie S1}: Trajectory showing no assembly without LLPS. Parameters are: $\Econd=0$, $\ebind=6$, $\rhoT=4\times10^{-4}$. 
    
    \item \textbf{Movie S2}: Capsid assembly trajectory without LLPS. Parameters are: $\Econd=0$, $\ebind=7$, $\rhoT=4\times10^{-4}$. 

    \item \textbf{Movie S3}: Capsid assembly trajectory with LLPS for relatively weak subunit partitioning. Parameters are: $\Econd=3$, $\ebind=6$, $\rhoT=4\times10^{-4}$, $\Vr=5.0\times10^{-3}$. 

    \item \textbf{Movie S4}: Capsid assembly trajectory with LLPS for relatively stong subunit partitioning. Parameters are: $\Econd=7$, $\ebind=6$, $\rhoT=4\times10^{-4}$, $\Vr=5.0\times10^{-3}$. 

    \item \textbf{Movie S5}: Capsid assembly trajectory with LLPS for a relatively large condensate volume fraction. Parameters are: $\Econd=7$, $\ebind=6$, $\rhoT=4\times10^{-4}$, $\Vr=0.11$. 

    \item \textbf{Movie S6}: Capsid assembly trajectory without LLPS at high subunit binding affinity, exhibiting malformed structures. Parameters are: $\Econd=0$, $\ebind=8$, $\rhoT=4\times10^{-4}$. 

    \item \textbf{Movie S7}: Capsid assembly trajectory with LLPS at high subunit binding affinity, exhibiting malformed structures. Parameters are: $\Econd=7$, $\ebind=8$, $\rhoT=4\times10^{-4}$, $\Vr=5.0\times10^{-3}$. 
\end{itemize}

\bibliography{references}